\documentclass[reprint, aps, prfluids, superscriptaddress, onecolumn, longbibliography, 12pt]{revtex4-2}

\usepackage{csquotes}
\usepackage[T1]{fontenc}
\usepackage[utf8]{inputenc}
\usepackage{amsmath}
\usepackage{amssymb}
\usepackage{enumitem}
\usepackage{graphicx}
\usepackage{refstyle}
\usepackage[
  citecolor=blue,
  colorlinks,
  linkcolor=blue,
  urlcolor=blue,
]{hyperref}
\usepackage[dvipsnames]{xcolor}
\begin{document}

%\preprint{}

\title{Flow structure, heat transport, and dissipation in vertical-field magnetoconvection at similar Reynolds numbers}

\author{Md Shahid Raza}
\affiliation{School of Mechanical and Materials Engineering, Indian Institute of Technology Mandi, Kamand 175075, India}

\author{Akanksha Bhardwaj}
\affiliation{School of Mechanical and Materials Engineering, Indian Institute of Technology Mandi, Kamand 175075, India}

\author{Shashwat Bhattacharya}
\email{shashwat@iitmandi.ac.in}
\affiliation{School of Mechanical and Materials Engineering, Indian Institute of Technology Mandi, Kamand 175075, India}

\date{\today}

\begin{abstract}
We employ direct numerical simulations to investigate the effect of a vertical magnetic field on the flow structures, heat transport, and dissipation rates in Rayleigh--B\'enard convection of a low-Prandtl-number fluid ($\mathrm{Pr} = 0.025$) at an approximately fixed Reynolds number ($\mathrm{Re} \approx 880$), maintained by adjusting the Rayleigh number to compensate for the Lorentz force. As the Hartmann number increases, the flow becomes more anisotropic, which strengthens the positive convective heat fluxes relative to the negative ones and thereby raises the Nusselt number, accompanied by a thinning of the thermal boundary layers. The total viscous and thermal dissipation rates are dominated by the boundary layers, whereas the total magnetic dissipation rate is dominated by the bulk. With increasing Hartmann number, the boundary-layer fraction of the viscous and magnetic dissipation rates increases, whereas that of the thermal dissipation computed from the temperature fluctuations decreases, owing to the suppression of temperature fluctuations near the boundary layers. The dissipation rates deviate from log-normality and are well described by stretched exponential distributions, whose tails become more extended with increasing Hartmann number.
\end{abstract}
%\keywords{}

\maketitle



\maketitle
\section{Introduction} \label{sec:Introduction}
Buoyancy-driven convection is ubiquitous across geophysical, astrophysical, and engineering systems. In electrically conducting media, external or self-generated magnetic fields couple to the flow, introducing the Lorentz force as a body force alongside buoyancy~\cite{Chandrasekhar:book:Instability,Weiss:book}. Such magnetoconvective flows occur in the Sun and planetary interiors~\cite{Davidson:book:MHD,Schumacher:RMP2020,Stefani:NRP2024}, as well as in technological applications such as magnetic stirring and braking~\cite{Davidson:ARFM1999,Lyubimov:JFM2010}, liquid-metal batteries~\cite{Kelley:PF2014,Shen:TCFD2016,Kelley:AMR2018,Davidson:PRF2022,Davidson:PRF2022}, and liquid-metal cooling blankets in next-generation fusion reactors~\cite{Buhler:FED2013,Mistrangelo:FED2020,Mistrangelo:FED2021}.

In the present work, we focus on magnetoconvection of liquid metals, in which density variations are negligible except in the buoyancy term of the momentum equation (the Boussinesq approximation)~\cite{Chandrasekhar:book:Instability}. Magnetoconvection is typically studied in the simplified setting of a fluid layer heated from below and cooled from above (Rayleigh--B\'enard convection) in the presence of a magnetic field. It is governed by four non-dimensional parameters: (i) the Rayleigh number $\mathrm{Ra}$, the ratio of buoyancy to dissipative forces; (ii) the Prandtl number $\mathrm{Pr}$, the ratio of the kinematic viscosity to the thermal diffusivity; (iii) the Hartmann number $\mathrm{Ha}$, the ratio of Lorentz to viscous forces; and (iv) the magnetic Prandtl number $\mathrm{Pm}$, the ratio of the kinematic viscosity to the magnetic diffusivity. The magnetic field strength is also commonly quantified by the magnetic interaction parameter $N$, the ratio of the Lorentz force to the inertial force.
The principal response parameters are the Nusselt number $\mathrm{Nu}$, the normalized heat flux; the Reynolds number $\mathrm{Re}$, which measures the momentum transport; and the magnetic Reynolds number $\mathrm{Re}_m$, the ratio of magnetic advection to magnetic diffusion. In liquid-metal convection in engineering applications and laboratory experiments, both $\mathrm{Re}_m$ and $\mathrm{Pm}$ are very small, so the induced magnetic field is negligible compared with the applied field. Such flows are called quasi-static magnetoconvection, in which the induced magnetic field adjusts instantaneously to changes in the velocity and is neglected in the expressions for the Lorentz force and Ohm's law~\citep{Roberts:book,Davidson:book:MHD,Verma:book:ET}.

The influence of magnetic fields on Rayleigh--B\'enard convection has been studied extensively using theory, experiments, and numerical simulations. An externally imposed horizontal magnetic field renders the flow quasi-two-dimensional, with the convection rolls aligning along the field direction~\cite{Fauve:JPL1981,Busse:JTAM1983,Vogt:PRF2018}, whereas a vertical magnetic field suppresses convective motion~\cite{Chandrasekhar:book:Instability,Cioni:JFM1997,King:PNAS2015}. In a Rayleigh--B\'enard system, convection sets in above a critical Rayleigh number, which, in the absence of a magnetic field, is $\mathrm{Ra}_c = 1708$ for no-slip horizontal walls~\cite{Chandrasekhar:book:Instability}. This critical value increases when a vertical magnetic field is applied, scaling as $\mathrm{Ra}_c \sim \mathrm{Ha}^2$ in the limit of large Hartmann numbers. Below $\mathrm{Ra}_c$, convection in the bulk ceases; however, in the presence of sidewalls, remnant convective modes localised near the sidewalls persist~\cite{Busse:PF2008,Liu:JFM2018,Zuerner:JFM2020,Akhmedagaev:JFM2020,Bhattacharya:JFM2023,Bhattacharya:JFM2024,McCormack:JFM2025,Tao:JFM2026}.
There are a few studies on convection with tilted magnetic fields; these studies report a tendency of the mean flow and wall modes to travel or align in the direction of the tilt~\cite{Hurlburt:ApJ1996,Bhattacharya:JFM2024}, as well as qualitative similarities to convection with a vertical magnetic field in terms of convection patterns, heat transport, and flow speed~\cite{Nicoski:PRF2022}.

Vertical magnetic fields can significantly alter the nature of turbulence, flow structures, and heat transport in Rayleigh--B\'enard convection, and this is the focus of the present paper. Simulations (e.g., Refs.~\cite{Yan:JFM2019,Akhmedagaev:JFM2020,McCormack:JFM2023,Teimurazov:JFM2024}) and experiments (e.g., Refs.~\cite{Zuerner:JFM2020,Xu:PRF2023}) of magnetoconvection with a vertical magnetic field have identified several regimes, distinguished by the strength of the magnetic field relative to the buoyant forcing. These regimes range from multicellular flows with strong anisotropy at strong magnetic fields to turbulent flows with a large-scale circulation at weak magnetic fields. Correspondingly, the scaling of the Nusselt number varies from $\mathrm{Nu} \sim \mathrm{Ra}$ at strong fields to the classical $\mathrm{Nu} \sim \mathrm{Ra}^{2/7}$ at weak fields~\cite{Cresswell:PRF2023}. Heat transport is suppressed at strong magnetic fields owing to the inhibition of turbulent motions, and has also been shown to fluctuate because of mode switching~\cite{Xu:PRF2023}.

It is well established that increasing the Hartmann number at a fixed Rayleigh number reduces the heat flux by suppressing turbulence. However, how the heat flux and flow structures vary with the Hartmann number at a \emph{fixed Reynolds number}, that is, at a fixed strength of turbulence, remains unclear. We address this gap by simulating turbulent magnetoconvection at several Hartmann numbers, adjusting the Rayleigh number so that the Reynolds number remains approximately constant, and examining the resulting flow structures and heat transport. In addition, we present what is, to our knowledge, the first comprehensive analysis of the statistics and spatial distribution of the thermal, viscous, and magnetic dissipation rates in magnetoconvection. Our observations are largely consistent with the physical arguments proposed in earlier studies.

The remainder of the paper is organized as follows. In Sec.~\ref{sec:Numerics}, we present the governing equations of quasi-static magnetoconvection and the simulation details. In Sec.~\ref{sec:Results}, we discuss the flow structures, temperature profiles, and heat flux for various magnetic field strengths, followed by the statistics of the viscous, thermal, and magnetic dissipation rates. We conclude in Sec.~\ref{sec:Conclusion}.

%\citet{Yan:JFM2019} classified magnetoconvection in an unbounded plane layer into three bulk-dominated regimes (cellular, columnar, and turbulent) based on internal force balances, whereas \citet{Akhmedagaev:JFM2020} and \citet{Zuerner:JFM2020} studied low-Prandtl-number fluids in bounded cylindrical containers ($\Gamma = 1$), where sidewall constraints fundamentally shape the flow patterns. At intermediate buoyancy forcing, the vertically aligned columnar state observed by \citet{Yan:JFM2019} contrasts with the quasi-two-dimensional planar vortex sheets of \citet{Akhmedagaev:JFM2020} and the multi-roll cellular patterns of \citet{Zuerner:JFM2020}. 
%Later, \citet{Xu:PRF2023}  a multicellular regime for intermediate buoyancy forcing in their experimental study of magnetoconvection in liquid gallium similar to \citet{Zuerner:JFM2020} and discussed fluctuations in the global heat transport due to mode switching. \citet{McCormack:JFM2023} exhibitted the coexistence of wall modes and large scale circulation at moderate Hartmann numbers. \citet{Cresswel:PRF2023} categorize strong-field Boussinesq magnetoconvection into magnetically constrained, transitional, and magnetically influenced regimes based on solenoidal force balances, demonstrating that heat transport transitions from $Nu \propto Ra$ to a classical $Nu \propto Ra^{2/7}$ scaling. 
%Under weak magnetic fields or extreme thermal driving, however, all the frameworks converge on a transition to turbulent flow.

\section{Numerical Methods} \label{sec:Numerics}
\noindent
\textbf{} We perform direct numerical simulations of convection under a vertically imposed magnetic field in a cubical domain of unit aspect ratio $(L_x= L_y=L_z=1)$. We employ the Boussinesq approximation in which the effects of changes in density are negligible except for the buoyancy term of the momentum equation. 
The study is conducted under the quasi-static approximation, where the induced magnetic field is negligible compared to the applied magnetic field; such approximation is appropriate for magnetoconvection in liquid metals. Under these assumptions, the governing equations for magnetoconvection are:
\begin{align}
	\boldsymbol{\nabla} \cdot \mathbf{u} &= 0 \label{eq:continuity_DNS} \\
	\frac{\partial \mathbf{u}}{\partial t} + \mathbf{u}\cdot \boldsymbol{\nabla} \mathbf{u} &= -\boldsymbol{\nabla} p + T\hat{\mathbf{z}} + \sqrt{\frac{\mathrm{Pr}}{\mathrm{Ra}}} \nabla^2 \mathbf{u} + \mathrm{Ha}_z^2\sqrt{\frac{\mathrm{Pr}}{\mathrm{Ra}}}(\mathbf{j} \times \hat{z}), \label{eq:Momentum_DNS} \\
	\frac{\partial T}{\partial t} + \mathbf{u} \cdot \boldsymbol{\nabla} T &= \frac{1}{\sqrt{\mathrm{Ra} \mathrm{Pr}}}\nabla^2 T. \label{eq:T_energy_DNS}
\end{align}

In the above, $\mathbf{u}$, $p$, $T$, $\mathbf{j}$, and $\hat{z}$ represent velocity, pressure, temperature, current density, and the unit vector along the $z$-direction, respectively.  
The parameters $\mathrm{Ra}$, $\mathrm{Pr}$, and $\mathrm{Ha}$ denote the Rayleigh, Prandtl, and Hartmann numbers, and are given by:
\begin{equation}
    \mathrm{Ra} = \frac{\alpha g \Delta d^3}{\nu \kappa}, \quad
    \mathrm{Pr} = \frac{\nu}{\kappa}, \quad
    \mathrm{Ha} = B d \sqrt{\frac{\sigma}{\rho \nu}},
\end{equation}
where $\alpha$ is the thermal expansion coefficient, $g$ is the gravitational acceleration, $d$ and $\Delta$ are, respectively, the distance between the top and bottom walls, $\nu$ and $\kappa$ are respectively the kinematic viscosity and thermal diffusivity of the fluid, $B$ is the magnitude of the externally imposed magnetic field, and $\sigma$ and $\rho$ are respectively the electrical conductivity and density of the fluid. Thus, the Rayleigh number is the measure of the strength of buoyancy forcing, and the Hartmann number is representative of the strength of the magnetic field and electrical conductivity of the fluid. The current density is given by the modified Ohm's law:
\begin{equation}
    \mathbf{j} = -{\boldsymbol{\nabla}} \phi + (\mathbf{u} \times {\hat{z}}),
\label{eq:Current_DNS}
\end{equation}
where $\phi$ is the electric potential. The current-density is divergence free, resulting in the following Poisson-equation for the electric potential:
\begin{equation}
\nabla^2 \phi = {\boldsymbol{\nabla}} \cdot (\mathbf{u} \times {\hat{z}}). 
\label{eq:Potential_DNS}
\end{equation}

We perform simulations of magnetoconvection in a cubic box for three cases using a finite difference solver developed by \citet{Krasnov:CAF2011,Krasnov:JCP2023}. The Prandtl number is fixed at $\mathrm{Pr} = 0.025$, typical of liquid metals. The Rayleigh and Hartmann numbers are varied so that the Reynolds number, based on the root-mean-square velocity $u_{rms}=\sqrt{\langle u_x^2+y_y^2+u_z^2\rangle_V}$ and the cell height $d$, remains approximately constant at $\mathrm{Re} = 880 \pm 20$ in all cases. In Case~I, no magnetic field is imposed and $\mathrm{Ra} = 7.5 \times 10^4$. In Case~II, we impose a weak magnetic field ($\mathrm{Ha} = 20$) and increase the Rayleigh number to $\mathrm{Ra} = 1 \times 10^5$. In Case~III, we impose a strong magnetic field ($\mathrm{Ha} = 220$) and increase the Rayleigh number further to $\mathrm{Ra} = 10^6$.

In all simulations, we employ no-slip boundary conditions on all walls. The top and bottom walls are isothermal, with $T = 0$ and $T = 1$, respectively, and the sidewalls are adiabatic. All walls are electrically insulating, corresponding to $\partial \phi / \partial \eta = 0$ for the electric potential $\phi$, where $\eta$ denotes the wall-normal coordinate. A tridiagonal scheme is used to solve the elliptic equations for the pressure and electric potential, as well as the implicitly treated diffusion term of the temperature equation. A constant time step between $5 \times 10^{-5}$ and $1 \times 10^{-4}$ free-fall time units is used, which satisfies the Courant--Friedrichs--Lewy (CFL) condition in all runs.

The grid size varies from $256^3$ to $384^3$ depending on the parameters. The grid is uniform for $\mathrm{Ha} = 0$ and $\mathrm{Ha} = 20$, and non-uniform for $\mathrm{Ha} = 220$, with stronger clustering of grid points near the walls. The grid size and clustering were chosen a priori according to the expected thickness of the velocity boundary layers. Since the measured Reynolds number $\mathrm{Re}$ is not known before a run, we estimate the thickness from the input parameters using the free-fall Reynolds number $\mathrm{Re}_f = \sqrt{\mathrm{Ra}/\mathrm{Pr}}$. In convection without a magnetic field, the boundary-layer thickness scales as $\delta \sim \mathrm{Re}_f^{-1/2}$. A magnetic field, however, significantly alters the thickness and profile of the boundary layers. At the top and bottom walls, which are perpendicular to the field, the velocity boundary layer becomes a Hartmann layer, whose thickness scales as $\delta \sim \mathrm{Ha}^{-1}$ at large Hartmann numbers. At the sidewalls, which are parallel to the field, it becomes a Shercliff layer, with $\delta \sim \mathrm{Ha}^{-1/2}$. To account for both fluid inertia and the magnetic field, we estimate the boundary-layer thickness as
\begin{equation}
    \delta = \min \left(\frac{1}{\sqrt{\mathrm{Re}_f}}, \frac{1}{\mathrm{Ha}} \right)
\end{equation}
for the top and bottom walls, and as
\begin{equation}
    \delta = \min \left(\frac{1}{\sqrt{\mathrm{Re}_f}}, \frac{1}{\sqrt{\mathrm{Ha}}} \right)
\end{equation}
for the sidewalls. We ensured that at least five grid points lie within the boundary layers and that the maximum grid spacing is smaller than the Kolmogorov length scale,
\begin{equation}
    \eta_K = \left( \frac{\nu^3}{\epsilon} \right)^{1/4} \sim \mathrm{Re}^{-3/4},
\end{equation}
where $\epsilon$ is the viscous dissipation rate. The simulations were run for 10 to 22 free-fall time units, with the velocity and temperature fields saved at every free-fall time unit. All reported quantities are averaged over these snapshots. The principal simulation parameters are listed in Table~\ref{table:SimulationParameters}.

In the next section, we will discuss the results.

\begin{table}[t]
\centering
\setlength{\tabcolsep}{10pt}
\caption{Simulation parameters for all cases at $\mathrm{Pr} = 0.025$. $\mathrm{Re}$ is the time-averaged Reynolds number with its standard deviation. $N_\mathrm{BL}$ is the number of grid points within the viscous boundary layer, taken as the minimum over the six walls. $\Delta x_{\max}$ is the maximum grid spacing and $\eta_K$ is the Kolmogorov length scale; the grids for Cases~I and II are uniform, so $\Delta x_{\max}$ equals the grid spacing. ``Snapshots'' is the number of saved fields used for averaging.}
\begin{tabular}{c c c c c c c c}
\hline\hline
Case & $\mathrm{Ha}$ & $\mathrm{Ra}$ & $\mathrm{Re}$ & Grid resolution & $N_\mathrm{BL}$ & $\Delta x_{\max}/\eta_K$ & Snapshots \\
\hline
I   & 0   & $7.5\times10^{4}$ & $870 \pm 10$ & $256^3$ & 6 & 0.623 & 22 \\
II  & 20  & $1\times10^{5}$   & $884 \pm 4$  & $256^3$ & 5 & 0.631 & 20 \\
III & 220 & $1\times10^{6}$   & $892 \pm 5$  & $384^3$ & 6 & 0.704 & 10 \\
\hline\hline
\end{tabular}
\label{table:SimulationParameters}
\end{table}

\section{Results and discussion} \label{sec:Results}
In this section, we present and analyse the results of our simulations.
We study the flow structures, temperature and velocity profiles, and heat fluxes along with their responses to the different magnetic field strengths and the buoyant forcing.

\subsection{Flow structures} \label{subsec:Flow_structures}
\begin{figure}[!t]
    \centering
    \includegraphics[scale=0.25]{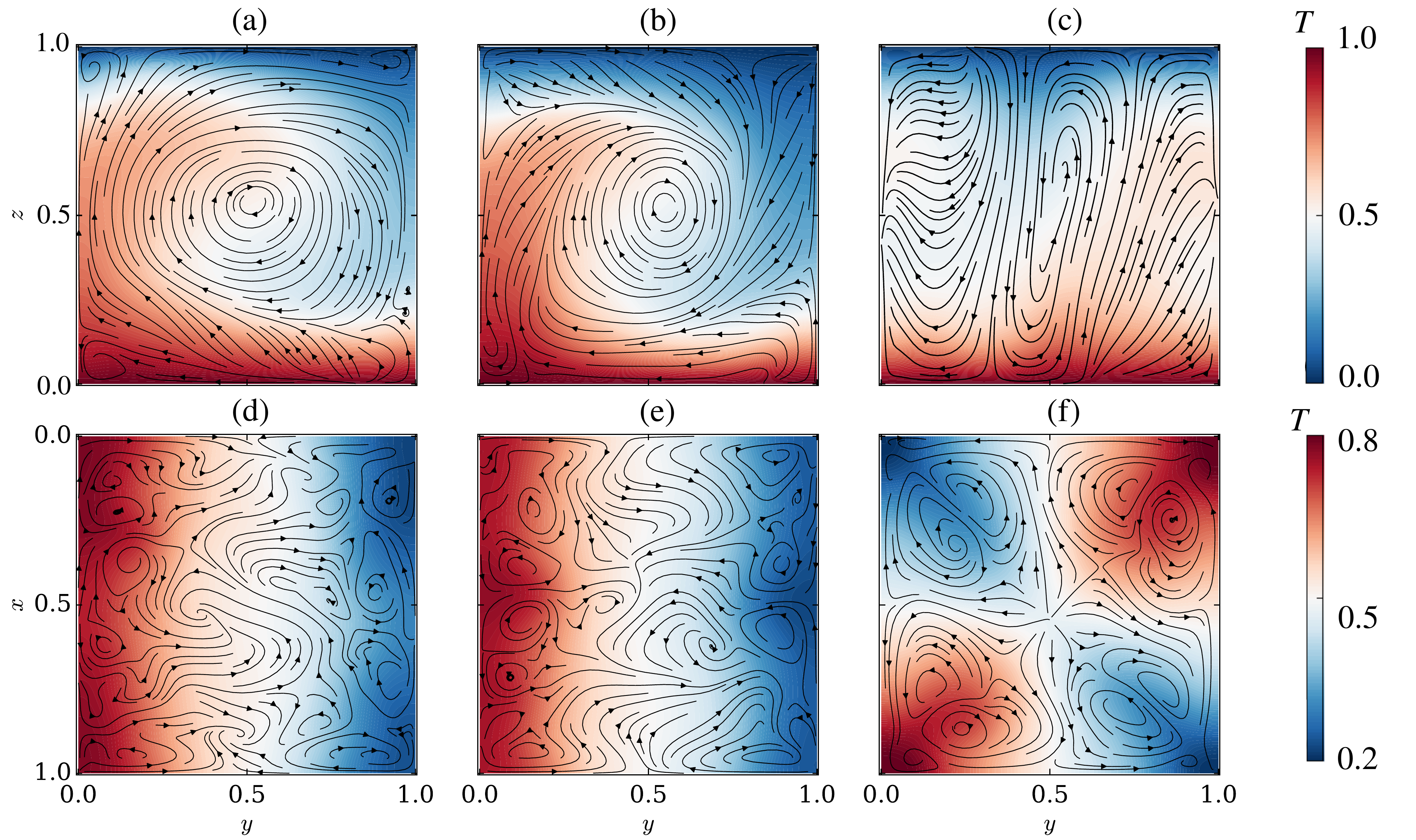}
    \caption{Contour plots of the temperature superimposed with velocity streamlines at vertical midplane corresponding to $x=0.5$ for (a) $\mathrm{Ha}=0$, $\mathrm{Ra}=7.5 \times 10^4$, (b) $\mathrm{Ha}=20$, $\mathrm{Ra}=1 \times 10^5$, and (c) $\mathrm{Ha}=220$, $\mathrm{Ra}=1 \times 10^6$, and at horizontal midplane for (d) $\mathrm{Ha}=0$, $\mathrm{Ra}=7.5 \times 10^4$, (e) $\mathrm{Ha}=20$, $\mathrm{Ra}=1 \times 10^5$, and (f) $\mathrm{Ha}=220$, $\mathrm{Ra}=1 \times 10^6$. Despite having similar Reynolds numbers, the flow structure with strong magnetic fields (corresponding to $\mathrm{Ha}=220$) is vastly different with columnar structures compared to the single roll structure at lower Hartmann numbers. }
    \label{fig:Contours}
\end{figure}
In this subsection, we analyze the flow structures across the three magnetoconvection configurations. In Figs.~\ref{fig:Contours}(a--c), we present the temperature contours overlaid with velocity streamlines on the vertical midplane ($x=0.5$). Similarly, Figs.~\ref{fig:Contours}(d--f) show the corresponding distributions on the horizontal midplane ($z=0.5$). 
Despite having similar Reynolds numbers (and hence comparable turbulence strengths), the three cases exhibit markedly distinct flow structures. This contrast is particularly pronounced in the third case, where there is a strong magnetic field ($\mathrm{Ha}=220$). For pure convection, shown in Fig.~\ref{fig:Contours}(a), the flow consists of a single large-scale circulation (LSC) roll accompanied by secondary corner rolls. The single-LSC structure is further confirmed by the distinct hot and cold plume bands adjacent to the $x=0$ and $x=d$ sidewalls in Fig.~\ref{fig:Contours}(b). As seen in Figs.~\ref{fig:Contours}(b) and (e), the LSC persists under a weak magnetic field ($\mathrm{Ha}=20$). However, the roll becomes noticeably more elliptical, with its major axis aligned along the vertical direction—indicating that the magnetic field tends to induce flow anisotropy. 
\begin{table}[t]
\centering
\setlength{\tabcolsep}{10pt}
\caption{For our magnetoconvection runs: Anisotropy parameter ($A$), the viscous boundary layer thickness ($\delta_u$), thermal boundary layer thickness ($\delta_T$), Nusselt number ($\mathrm{Nu}$), and the ratio of the positive to negative heat fluxes ($ |\sum u_z T'^+/\sum u_z T'^-|$). }

\begin{tabular}{c c c c c c c}
\hline\hline
Case &  $N$ & $A$ & $\delta_u$ & $\delta_T$ & $\mathrm{Nu}$ & $ |\sum u_z T'^+/\sum u_z T'^-|$ \\
\hline
I ($\mathrm{Ha}=0$, $\mathrm{Ra}=7.5\times10^{4}$) & 0 & 0.53& 0.0242 & 0.174 & 2.87 & 9.15 \\
II ($\mathrm{Ha}=20$, $\mathrm{Ra}=1\times10^{5}$) & 0.45 & 0.52& 0.0223 & 0.168 & 2.97 & 10.7 \\
III ($\mathrm{Ha}=220$, $\mathrm{Ra}=1\times10^{6}$) & 54 & 0.29& 0.0047 &  0.140 & 3.57 & 16.0\\
\hline\hline
\end{tabular} \label{table:Nu_A}
\end{table} 

For the case with $\mathrm{Ha}=220$ (strong magnetic field), the LSC roll disintegrates into multiple convection rolls elongated along the vertical direction, as shown in Figs.~\ref{fig:Contours}(c) and (f). This disintegration has been observed in previous studies and is suggested to happen due to magnetic damping~\cite{McCormack:JFM2023}. Specifically, Fig.~\ref{fig:Contours}(f) reveals the presence of two parallel rolls situated adjacent to the $y=0$ and $y=1$ boundaries, respectively. These distributions demonstrate pronounced flow anisotropy characterized by enhanced vertical velocity fluctuations. Our observations align well with previous experimental~\cite{Zuerner:JFM2020} and numerical~\cite{Akhmedagaev:JFM2020} studies of magnetoconvection under strong magnetic fields.

To quantify the flow anisotropy across the three cases, we compute the anisotropy parameter $A$, defined as~\cite{Nath:PRF2016}:
\begin{equation}
    A = \frac{E_\perp}{2E_\parallel}=\frac{\langle u_x^2 + u_y^2 \rangle}{2 \langle u_z^2 \rangle},
    \label{eq:Anisotropy_parameter}
\end{equation}
where $E_\perp$ and $E_\parallel$ denote the kinetic energies associated with the velocity components perpendicular and parallel to the direction of buoyancy, respectively, and $\langle \cdot \rangle$ represents volume average. According to Eq.~(\ref{eq:Anisotropy_parameter}), a fully isotropic flow corresponds to $A = 1$. 
The computed anisotropy parameters are summarized in Table~\ref{table:Nu_A}. In all cases, $A$ is less than unity, implying that $\langle u_x^2 \rangle \approx \langle u_y^2 \rangle < \langle u_z^2 \rangle$. This behavior is expected, as buoyancy preferentially accelerates the flow along the $z$-direction. Notably, Table~\ref{table:Nu_A} shows that $A$ decreases further with an increasing Hartmann number. This reduction indicates an enhancement of the vertical flow, corroborating the velocity field visualizations presented in Fig.~\ref{fig:Contours}.

Next, we examine the vertical profiles of the horizontally averaged horizontal and vertical root-mean-square (rms) velocities, denoted as $u_\perp(z)$ and $u_\parallel(z)$, respectively. These quantities are defined as $u_{\perp}(z) = \sqrt{2 \langle E_\perp \rangle_{x,y}}$ and $u_{\parallel}(z) = \sqrt{2 \langle E_\parallel \rangle_{x,y}}$, where $\langle \cdot \rangle_{x,y}$ denotes planar averaging over the $x$--$y$ plane.
\begin{figure}[!t]
    \centering
    \includegraphics[scale=0.24]{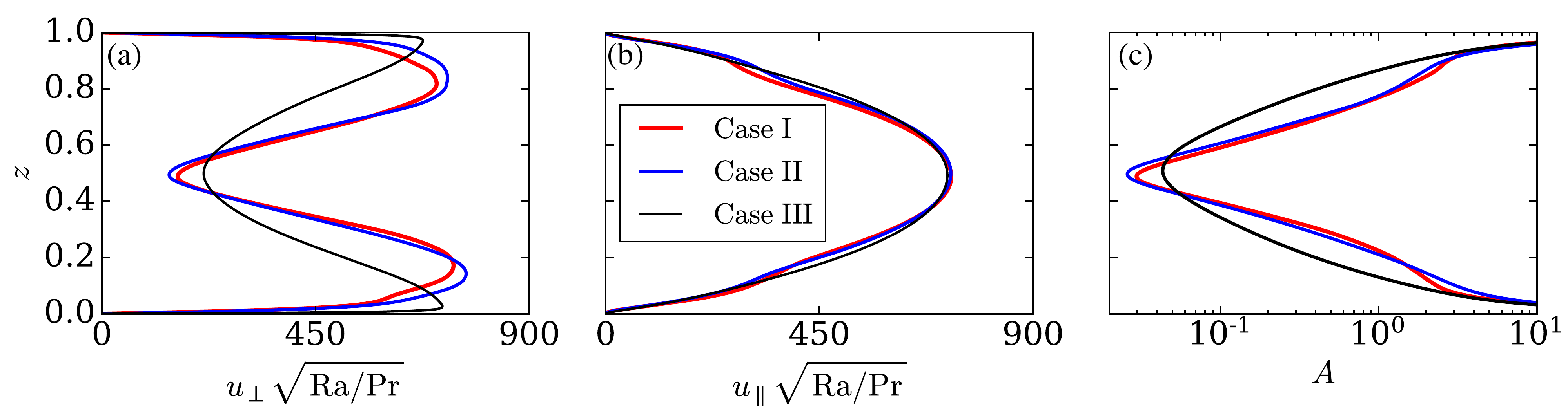}
    \caption{Profiles of (a) horizontal and (b) vertical root mean square velocities, averaged over horizontal every plane, along $z$. (c) The profile of anisotropy parameter, averaged over every horizontal plane, along $z$. While the vertical velocity profiles are similar for all our cases, the horizontal profile exhibits steeper gradients near the walls for Case III (convection with strong magnetic fields). Moreover, the anisotropy parameter is generally lower for the case with strong magnetic fields, except near the cell center.}
    \label{fig:E_A}
\end{figure}
Figures~\ref{fig:E_A}(a) and (b) present the vertical profiles of $u_\perp(z)$ and $u_\parallel(z)$, respectively, both normalized by $\sqrt{\mathrm{Pr}/\mathrm{Ra}}$. Compared to Cases I and II, the normalized $u_\perp(z)$ in Case III rises significantly more steeply near the horizontal boundaries, indicating thinner viscous boundary layers. This thinning is attributed to the influence of the magnetic field, which induces the formation of Hartmann layers. Furthermore, the normalized horizontal velocity in Case III is generally suppressed relative to Cases I and II throughout most of the domain, except near the cell center. In contrast, the normalized $u_\parallel(z)$ profiles remain largely similar across all three cases. These trends indicate that the magnetic field predominantly affects the horizontal velocity component while leaving the vertical motion largely unaltered.

Figure~\ref{fig:E_A}(c) shows the vertical distribution of the horizontally averaged anisotropy parameter. For the strong magnetic field case (Case III), the anisotropy parameter is consistently lower than in the other cases across most of the domain—except near the cell center. This further confirms that the imposed magnetic field suppresses the fluctuations along the horizontal direction due to Joule dissipation. The Joule dissipation is, in turn, governed by the magnetic interaction parameter, $N$, which represents the ratio of the Lorentz force to the inertial force, and is given by:
\begin{equation}
    N = \frac{\mathrm{Ha}^2}{\mathrm{Re}}.
    \label{eq:N}
\end{equation}
The calculated values of $N$ for our cases are listed in Table~\ref{table:Nu_A}. The first two configurations possess zero and very low magnetic interaction parameters, respectively ($N < 1$). In contrast, for the third case, the magnetic interaction parameter is substantially larger than unity ($N = 54 \gg 1$). This dominant Lorentz force accounts for the marked increase in anisotropy and the distinct flow structures observed in the third case compared to the other two.

To further characterize the near-wall flow, we extract the viscous boundary layer thickness, $\delta_u$, from our numerical data. Here, $\delta_u$ is defined as the wall-normal distance at which the linear slope of $u_\perp(z)$ at the wall intersects the tangent at its near-wall local maximum. The computed values of $\delta_u$, summarized in Table~\ref{table:Nu_A}, demonstrate that the boundary layer thins with an increasing Hartmann number. This trend directly corroborates the steepened $u_\perp(z)$ profiles observed in Fig.~\ref{fig:E_A}(a).
\begin{figure}[!t]
    \centering
    \includegraphics[scale=0.6]{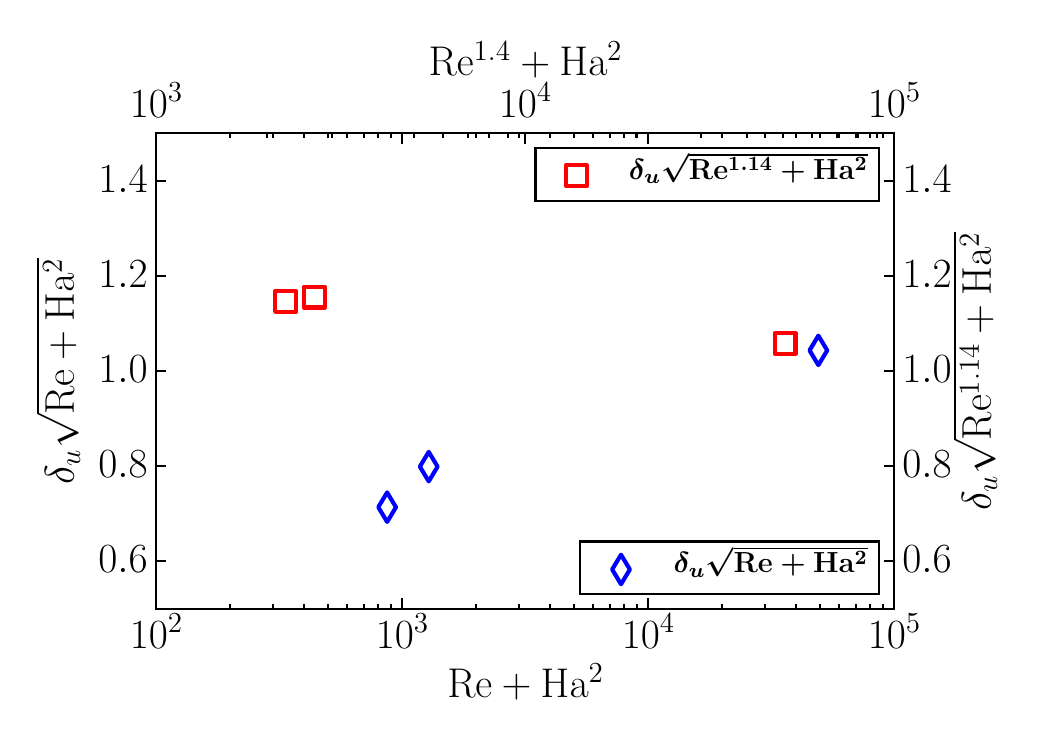}
    \caption{Compensated viscous boundary-layer thickness $\delta_u$ for the three cases: $\delta_u\sqrt{\mathrm{Re}+\mathrm{Ha}^2}$ versus $\mathrm{Re}+\mathrm{Ha}^2$ (blue diamonds), and the modified compensation $\delta_u\sqrt{\mathrm{Re}^{1.14}+\mathrm{Ha}^2}$ versus $\mathrm{Re}^{1.14}+\mathrm{Ha}^2$ (red squares). The modified compensated thickness is flatter, indicating that Eq.~(\ref{eq:BL_ReHa_Mod}) describes the boundary-layer thickness better than Eq.~(\ref{eq:BL_ReHa}).
    }
    \label{fig:BL}
\end{figure}

For pure convection at large Reynolds numbers with a laminar boundary layer, it is generally assumed that the viscous boundary layer thickness scales as $\delta_u \sim \mathrm{Re}^{-1/2}$, analogous to the flow over a flat plate. In magnetohydrodynamic (MHD) flows, $\delta_u$ also depends on the Hartmann number, scaling as $\delta_u \sim \mathrm{Ha}^{-1}$ under strong magnetic fields ($\mathrm{Ha}^2 \gg \mathrm{Re}$). In regimes where both $\mathrm{Re}$ and $\mathrm{Ha}$ are significant, dimensional analysis of the momentum equation yields the following scaling relation for $\delta_u$~\cite{Lim:JFM2019,Bhattacharya:JFM2023}:
\begin{equation}
    \delta_u \sim \frac{1}{\sqrt{\mathrm{Re} + \mathrm{Ha}^2}}.
    \label{eq:BL_ReHa}
\end{equation}
To verify this scaling, we plot the compensated viscous boundary layer thickness, $\delta_u \sqrt{\mathrm{Re} + \mathrm{Ha}^2}$, against $\mathrm{Re} + \mathrm{Ha}^2$ in Fig.~\ref{fig:BL}. The compensated thickness exhibits a marginal increase with $\mathrm{Re} + \mathrm{Ha}^2$, indicating that the ideal relation provided by Eq.~(\ref{eq:BL_ReHa}) does not strictly hold. Indeed, previous studies~\cite{Breuer:PRE2004,Scheel:JFM2012,Shi:JFM2012,Bhattacharya:PF2020} have established that even in pure convection, $\delta_u$ deviates from the exact $\mathrm{Re}^{-1/2}$ scaling. This deviation arises because the boundary layers in thermal convection behave differently than classical flat-plate flows, primarily due to the complex near-wall dynamics driven by erupting thermal plumes.

In a numerical study of pure convection covering a wide range of Rayleigh and Prandtl numbers, \citet{Bhattacharya:PF2020} demonstrated that the scaling of the viscous boundary layer thickness ranges from $\delta_u \sim \mathrm{Re}^{-0.2}$ in the high-$\mathrm{Pr}$ regime to $\delta_u \sim \mathrm{Re}^{-0.6}$ in the low-$\mathrm{Pr}$ regime. Specifically, at $\mathrm{Pr}=0.02$, $\delta_u$ was observed to scale as $\delta_u \sim \mathrm{Re}^{-0.57}$. Given that our current cases operate at a comparable Prandtl number ($\mathrm{Pr}=0.025$), it is reasonable to assume this $\mathrm{Re}$-dependence remains valid. Consequently, for convection in the presence of magnetic fields, Eq.~(\ref{eq:BL_ReHa}) can be modified to:
\begin{equation}
    \delta_u \sim \frac{1}{\sqrt{\mathrm{Re}^{1.14} + \mathrm{Ha}^2}}.
    \label{eq:BL_ReHa_Mod}
\end{equation}
The modified compensated boundary layer thickness, $\delta_u \sqrt{\mathrm{Re}^{1.14} + \mathrm{Ha}^2}$, is plotted against $\mathrm{Re}^{1.14} + \mathrm{Ha}^2$ in Fig.~\ref{fig:BL}. As evident from the figure, this new compensated profile (shown in red) is significantly flatter than the original scaling curve (shown in blue). Quantitatively, the modified compensated values vary by only 8\%, compared to a 32\% variation observed with the original compensated thickness. Thus, our observations further corroborate the findings of \citet{Bhattacharya:PF2020}.

Our findings demonstrate that even at comparable levels of turbulence, the magnetic field significantly alters the underlying flow structures. In the next subsection, we will discuss the temperature profiles and the heat flux characteristics.

\subsection{Temperature profile and heat flux} \label{sec:HeatFlux}

\begin{figure}[!t]
    \centering
    \includegraphics[scale=0.24]{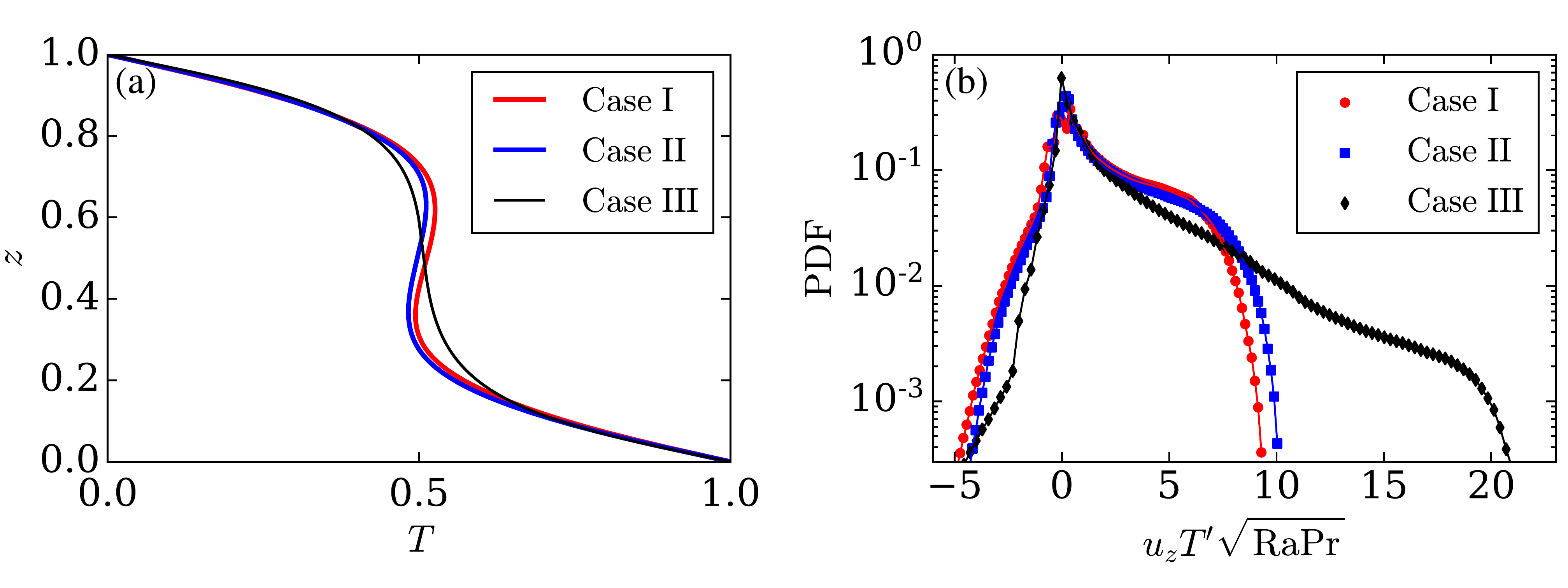}
    \caption{For our runs: (a) Profiles of temperature, averaged over every horizontal plane, versus $z$. The temperature is more well-mixed in the bulk for cases with weak or no magnetic fields. (b) Probability density functions (PDF) of convective heat flux normalised by the dimensionless thermal diffusivity $1/\sqrt{\mathrm{Ra} \, \mathrm{Pr}}$. The right tail is more extended for the case with strong magnetic field (Case III), implying higher net convective heat flux.}
    \label{fig:TempProfile}
\end{figure}

We now examine the impact of the magnetic field on the heat flux and temperature distribution within the domain. Figure~\ref{fig:TempProfile}(a) presents the horizontally averaged temperature profiles along the vertical direction. For Cases I and II, where the magnetic field is either absent ($\mathrm{Ha}=0$) or weak ($\mathrm{Ha}=20$), the bulk temperature is significantly more uniform than in Case III, which features a strong magnetic field ($\mathrm{Ha}=220$). Specifically, Case III exhibits a noticeable negative temperature gradient within the bulk region. In contrast, Cases I and II display a marginally positive temperature gradient in the bulk. The reduced thermal mixing observed in Case III is likely attributable to the suppression of horizontal velocity fluctuations by the strong magnetic field.

We now compute the thermal boundary-layer thickness $\delta_T$ for the three cases, defined as the wall-normal distance at which the linear extrapolation of the near-wall temperature profile intersects the mean bulk temperature, $T_\mathrm{bulk} = 0.5$. The computed values, listed in Table~\ref{table:Nu_A}, show that the thermal boundary layer thins with increasing Hartmann number. This thinning implies an enhanced heat flux, since the Nusselt number is related to $\delta_T$ by $\mathrm{Nu} \approx 1/(2\delta_T)$.

The Nusselt number, the normalized total heat flux, is defined as
\begin{equation}
    \mathrm{Nu} = 1 + \sqrt{\mathrm{Ra}\,\mathrm{Pr}}\,\langle u_z T' \rangle,
    \label{eq:Nu}
\end{equation}
where $T' = T - T_\mathrm{bulk}$. The computed Nusselt numbers, also listed in Table~\ref{table:Nu_A}, confirm that $\mathrm{Nu}$ increases with $\mathrm{Ha}$, from $\mathrm{Nu} = 2.87$ in Case~I to $\mathrm{Nu} = 3.57$ in Case~III. At first glance, this increase might be attributed to the slight rise in the Reynolds number across the cases, from $870$ to $892$. To test this, we use the classical non-magnetic scalings at fixed $\mathrm{Pr}$, $\mathrm{Nu} \sim \mathrm{Ra}^{2/7}$ and $\mathrm{Re} \sim \mathrm{Ra}^{1/2}$, as a reference for the behaviour expected in the absence of magnetic effects. Together they give $\mathrm{Nu} \sim \mathrm{Re}^{4/7}$, which predicts an increase of only about $1\%$, from $2.87$ to $2.91$. The measured increase of about $24\%$ is far larger, so the change in the Reynolds number cannot explain it. This behaviour is counterintuitive: despite the suppression of turbulent mixing by the magnetic field, the overall heat transport across the domain increases.

We further explore the statistics of the convective heat flux to understand the dynamics of heat transfer under varying magnetic fields. In buoyancy-driven convection, since hot plumes ascend and cold plumes descend, $u_z$ and $T'$ are positively correlated; hence, $\langle u_z T' \rangle > 0$~\cite{Verma:NJP2017}. Consequently, the convective contribution to the Nusselt number is strictly positive. However, the local convective heat flux, $u_z T'$, can assume both positive and negative values locally, despite having a positive global mean. The statistical distribution of these local fluxes directly impacts the overall Nusselt number.

To analyze these statistics, we compute the probability density functions (PDFs) of the local convective heat flux, normalized by the dimensionless thermal diffusivity ($1/\sqrt{\mathrm{Ra} \mathrm{Pr}}$), and plot them in Fig.~\ref{fig:TempProfile}(b). Across all cases, the PDFs exhibit a positive skew with pronounced right tails, reflecting the positive mean of $u_z T$. Furthermore, the right tail extends significantly as the Hartmann number increases; this broadening is particularly prominent under the strong magnetic field in Case III. This indicates that the magnitude of positive heat flux events increases with the magnetic field, consistent with our earlier observations of an enhanced global Nusselt number.

To quantify the relative contributions of the positive and negative fluxes, we define the ratio $R = \left| \frac{\sum (u_z T')^+}{\sum (u_z T')^-} \right|$, where $(u_z T')^+$ and $(u_z T')^-$ denote the positive and negative local values of $u_z T'$, respectively~\cite{Tiwari:PNAS2025}. As listed in Table~\ref{table:Nu_A}, $R$ exhibits an upward trend with an increasing Hartmann number, rising from $R \approx 9$ for Case I ($\mathrm{Ha}=0$) to $R \approx 16$ for Case III ($\mathrm{Ha}=220$). This trend implies that the net heat flux enhancement is primarily driven by an increasing dominance of positive heat fluxes over negative ones, rather than by enhanced bulk mixing (which is typically the primary heat transfer mechanism in pure turbulent convection). Thus, although strong magnetic fields suppress bulk mixing, the resulting flow anisotropy significantly amplifies the relative contribution of the positive convective heat flux.

In the next subsection, we will discuss the statistics of viscous, thermal, and magnetic dissipation rates.

\subsection{Dissipation rate statistics} \label{sec:Dissipation}
Viscous, thermal, and magnetic dissipation rates play a central role in global heat transport within buoyancy-driven convection. These quantities represent the direct dissipation of kinetic, thermal, and magnetic energies governed by kinematic viscosity, thermal diffusivity, and magnetic diffusivity, respectively. The dimensionless local dissipation rates are defined as
\begin{align}
    \epsilon_u (\mathbf{r}) &= \frac{1}{2}\sqrt{\frac{\mathrm{Pr}}{\mathrm{Ra}}}(\partial_i u_j + \partial_j u_i)^2, \label{eq:VD} \\
    \epsilon_b (\mathbf{r}) &= \mathrm{Ha}^2 \sqrt{\frac{\mathrm{Pr}}{\mathrm{Ra}}} j_i j_i, \label{eq:MD} \\
    \epsilon_T (\mathbf{r}) &= \frac{1}{\sqrt{\mathrm{Ra}\,\mathrm{Pr}}} (\nabla T)^2. \label{eq:TD}
\end{align}
Here, $\epsilon_u (\mathbf{r})$, $\epsilon_b (\mathbf{r})$, and $\epsilon_T (\mathbf{r})$ denote the local viscous, magnetic, and thermal dissipation rates at coordinate vector $\mathbf{r}$. Their volume-averaged dimensional counterparts are directly linked to the Nusselt number via~\cite{Shraiman:PRA1990,Zurner:PRE2016,Zurner:PF2020}:
\begin{equation}
    \langle \epsilon_u \rangle + \langle \epsilon_b \rangle = \frac{\nu^3}{d^4}(\mathrm{Nu}-1)\frac{\mathrm{Ra}}{\mathrm{Pr}^2},  \quad
    \langle \epsilon_T \rangle = \frac{\kappa \Delta^2}{d^2} \mathrm{Nu}, \label{eq:SS}
\end{equation}
where $\Delta$ in Eq.~(\ref{eq:SS}) represents the temperature difference across the driving boundaries.

Understanding the scaling and spatial statistics of dissipation rates across the bulk and boundary layers is essential for developing unified models of thermal convection~\cite{Grossmann:PRL2001,Zurner:PF2020,Bhattacharya:PF2020}. The total dissipation rates ($\tilde{D}_u = \langle \epsilon_u \rangle V$, $\tilde{D}_T = \langle \epsilon_T \rangle V$, and $\tilde{D}_b = \langle \epsilon_b \rangle V$, where $V$ is the domain volume) decompose into bulk and boundary-layer contributions:
\begin{align}
    \tilde{D}_u &= \int_{\mathrm{bulk}} \epsilon_u (\mathbf{r}) \, \mathrm{d}^3 \mathbf{r} + \int_{\mathrm{BL}} \epsilon_u (\mathbf{r}) \, \mathrm{d}^3 \mathbf{r} \nonumber \\
               &= \tilde{D}_{u,\mathrm{bulk}} + \tilde{D}_{u,\mathrm{BL}}, \label{eq:VDiss_BulkBL} \\[6pt]
    \tilde{D}_T &= \int_{\mathrm{bulk}} \epsilon_T (\mathbf{r}) \, \mathrm{d}^3 \mathbf{r} + \int_{\mathrm{BL}} \epsilon_T (\mathbf{r}) \, \mathrm{d}^3 \mathbf{r} \nonumber \\
               &= \tilde{D}_{T,\mathrm{bulk}} + \tilde{D}_{T,\mathrm{BL}}, \label{eq:TDiss_BulkBL} \\[6pt]
    \tilde{D}_b &= \int_{\mathrm{bulk}} \epsilon_b (\mathbf{r}) \, \mathrm{d}^3 \mathbf{r} + \int_{\mathrm{BL}} \epsilon_b (\mathbf{r}) \, \mathrm{d}^3 \mathbf{r} \nonumber \\
               &= \tilde{D}_{b,\mathrm{bulk}} + \tilde{D}_{b,\mathrm{BL}}. \label{eq:MDiss_BulkBL}
\end{align}
where the subscripts ``bulk'' and ``BL'' denote the bulk and boundary-layer regions, respectively. Which of these two regions dominates the global dissipation determines how the Nusselt number scales with the governing control parameters.

We numerically compute the dissipation rates from our numerical data and calculate the relative contributions of the bulk and boundary layers to the total dissipation rate, as listed in Table~\ref{table:Dissipation}.
For weak or absent magnetic fields, the viscous dissipation rates in the bulk and boundary layers are of the same order, with the boundary layer contribution marginally exceeding that of the bulk. This contrasts with $\mathrm{Pr} \sim 1$ convection, where bulk dissipation dominates~\cite{Bhattacharya:PF2018,Bhattacharya:PF2020}. For strong magnetic fields, the boundary layer contribution becomes increasingly dominant, reaching up to three times that of the bulk. 
%This is despite the fact that the viscous boundary layers become thinner with the increase in the artmann number, implying that the intensity of the dissi 

The thermal dissipation rate is likewise boundary-layer dominated (Table~\ref{table:Dissipation}), with the boundary-layer contribution several times larger than that of the bulk for weak or absent magnetic fields, consistent with earlier studies~\cite{Bhattacharya:PF2019,Bhattacharya:PF2020}. This is largely attributable to the steep temperature gradients within the thermal boundary layers, where the temperature transitions sharply from its wall value to that of the bulk. Notably, however, the boundary-layer contribution weakens considerably in Case~III, where the magnetic field is strong --- despite the persistence of sharp near-wall temperature gradients and a thermal boundary layer thinner (albeit marginally) than in the other two cases.
\begin{table}[t]
\centering
\setlength{\tabcolsep}{10pt}
\caption{Boundary-layer-to-bulk ratios of the total and volume-averaged dissipation rates for all three magnetoconvection cases. The total dissipation ratios are shown for viscous ($\tilde{D}_{u,\mathrm{BL}}/\tilde{D}_{u,\mathrm{bulk}}$), thermal ($\tilde{D}_{T,\mathrm{BL}}/\tilde{D}_{T,\mathrm{bulk}}$ and $\tilde{D}_{\theta,\mathrm{BL}}/\tilde{D}_{\theta,\mathrm{bulk}}$), and magnetic ($\tilde{D}_{b,\mathrm{BL}}/\tilde{D}_{b,\mathrm{bulk}}$) dissipation; the volume-averaged ratios are shown for viscous ($\epsilon_{u,\mathrm{BL}}/\epsilon_{u,\mathrm{bulk}}$), thermal ($\epsilon_{\theta,\mathrm{BL}}/\epsilon_{\theta,\mathrm{bulk}}$), and magnetic ($\epsilon_{b,\mathrm{BL}}/\epsilon_{b,\mathrm{bulk}}$) dissipation. Dashes indicate cases where the magnetic field is absent.}
\begin{tabular}{c c c c c c c c}
\hline\hline
Case &  $\frac{\tilde{D}_{u,\mathrm{BL}}}{\tilde{D}_{u,\mathrm{bulk}}}$ & $\frac{\tilde{D}_{T,\mathrm{BL}}}{\tilde{D}_{T,\mathrm{bulk}}}$ &
$\frac{\tilde{D}_{\theta,\mathrm{BL}}}{\tilde{D}_{\theta,\mathrm{bulk}}}$ &
$\frac{\tilde{D}_{b,\mathrm{BL}}}{\tilde{D}_{b,\mathrm{bulk}}}$ & $\frac{{\epsilon}_{u,\mathrm{BL}}}{{\epsilon}_{u,\mathrm{bulk}}}$ & $\frac{{\epsilon}_{\theta,\mathrm{BL}}}{{\epsilon}_{\theta,\mathrm{bulk}}}$ & $\frac{{\epsilon}_{b,\mathrm{BL}}}{{\epsilon}_{b,\mathrm{bulk}}}$ \\
\hline
I ($\mathrm{Ha}=0$) & 1.25 & 4.61& 1.53 & -  & 9.02 & 2.86 & - \\
II ($\mathrm{Ha}=20$) & 1.57 & 4.70& 1.49 &  0.133 & 11.8 & 2.94 & 0.997 \\
III ($\mathrm{Ha}=220$) & 3.29 & 1.13& 0.69 &  0.147 & 34.8 & 1.76 & 1.55 \\
\hline\hline
\end{tabular} \label{table:Dissipation}
\end{table} 
To investigate this anomaly further, we decompose the temperature field as
\begin{equation}
    T(x,y,z) = \bar{T}(z) + \theta(x,y,z), \label{eq:TDecomposition}
\end{equation}
where $\bar{T}(z)$ is the horizontally averaged temperature at height $z$, and $\theta$ is the temperature fluctuation about this mean. Subtracting the mean field removes the background inhomogeneity due to thermal stratification and the steep mean temperature gradient in the boundary layer, thereby isolating the contribution of pure turbulent mixing.
We define the thermal dissipation rate based on the temperature fluctuation $\theta$ as
\begin{equation}
    \tilde{D}_\theta = \int_V \kappa (\nabla \theta)^2 \, \mathrm{d}V.
\end{equation}
Since the temperature fluctuations $\theta$ have zero mean on every horizontal plane, the cross term vanishes upon integration, and the total thermal dissipation rate decomposes as
\begin{equation}
    \tilde{D}_T = \int_V \kappa \left(\nabla \bar{T}\right)^2 \mathrm{d}V 
    + \int_V \kappa \left(\nabla \theta\right)^2 \mathrm{d}V
    = \tilde{D}_{\bar{T}} + \tilde{D}_\theta.
\end{equation}
Thus, the total thermal dissipation rate comprises contributions from the mean temperature profile and the temperature fluctuations. Since the Nusselt number and mean temperature profile vary only modestly across the three cases (see Sec.~\ref{sec:HeatFlux}), the contribution of $\tilde{D}_{\bar{T}}$ to the boundary-layer thermal dissipation is expected to be similar in all three cases. It follows that the decrease in $\tilde{D}_{T,\mathrm{BL}}/\tilde{D}_{T,\mathrm{bulk}}$ in Case~III must be driven primarily by a decrease in $\tilde{D}_{\theta,\mathrm{BL}}/\tilde{D}_{\theta,\mathrm{bulk}}$.

To verify this, we compute the bulk and boundary-layer contributions to $\tilde{D}_\theta$ for all three cases and list them in Table~\ref{table:Dissipation}. As expected, the boundary-layer contribution to $\tilde{D}_\theta$ is smaller than that to $\tilde{D}_T$, owing to the removal of the $\tilde{D}_{\bar{T}}$ contribution. For weak or absent magnetic fields, while $\tilde{D}_{T,\mathrm{BL}}$ exceeds $\tilde{D}_{T,\mathrm{bulk}}$ by more than a factor of four, $\tilde{D}_{\theta,\mathrm{BL}}$ exceeds $\tilde{D}_{\theta,\mathrm{bulk}}$ by only a factor of 1.5. Nevertheless, since $\tilde{D}_{\theta,\mathrm{BL}} > \tilde{D}_{\theta,\mathrm{bulk}}$, temperature fluctuations remain stronger in the boundary layer than in the bulk.
For strong magnetic fields ($\mathrm{Ha} = 220$), $\tilde{D}_{\theta,\mathrm{BL}}/\tilde{D}_{\theta,\mathrm{bulk}}$ drops to $0.69$, indicating that $\tilde{D}_\theta$ is now bulk-dominated. Thus, the amplitude of temperature fluctuations in the boundary layer decreases relative to the bulk as the magnetic field strength increases. Consistent with our earlier conjecture, this reduction in $\tilde{D}_{\theta,\mathrm{BL}}/\tilde{D}_{\theta,\mathrm{bulk}}$ suppresses the total thermal dissipation in the boundary layer at large Hartmann numbers, despite the persistence of a steep background temperature gradient.

Finally, we examine the magnetic dissipation rates for Cases~II and III, with values of $\tilde{D}_{b,\mathrm{BL}}/\tilde{D}_{b,\mathrm{bulk}}$ listed in Table~\ref{table:Dissipation}. The magnetic dissipation is strongly bulk-dominated, as expected: since magnetic dissipation depends on the magnitude of the electric currents, the electrically insulating walls suppress near-wall currents and thereby weaken dissipation in the boundary layers. In both Cases~II and III, the bulk magnetic dissipation is nearly ten times larger than that in the boundary layers.

\begin{figure}[!t]
    \centering
    \includegraphics[scale=0.25]{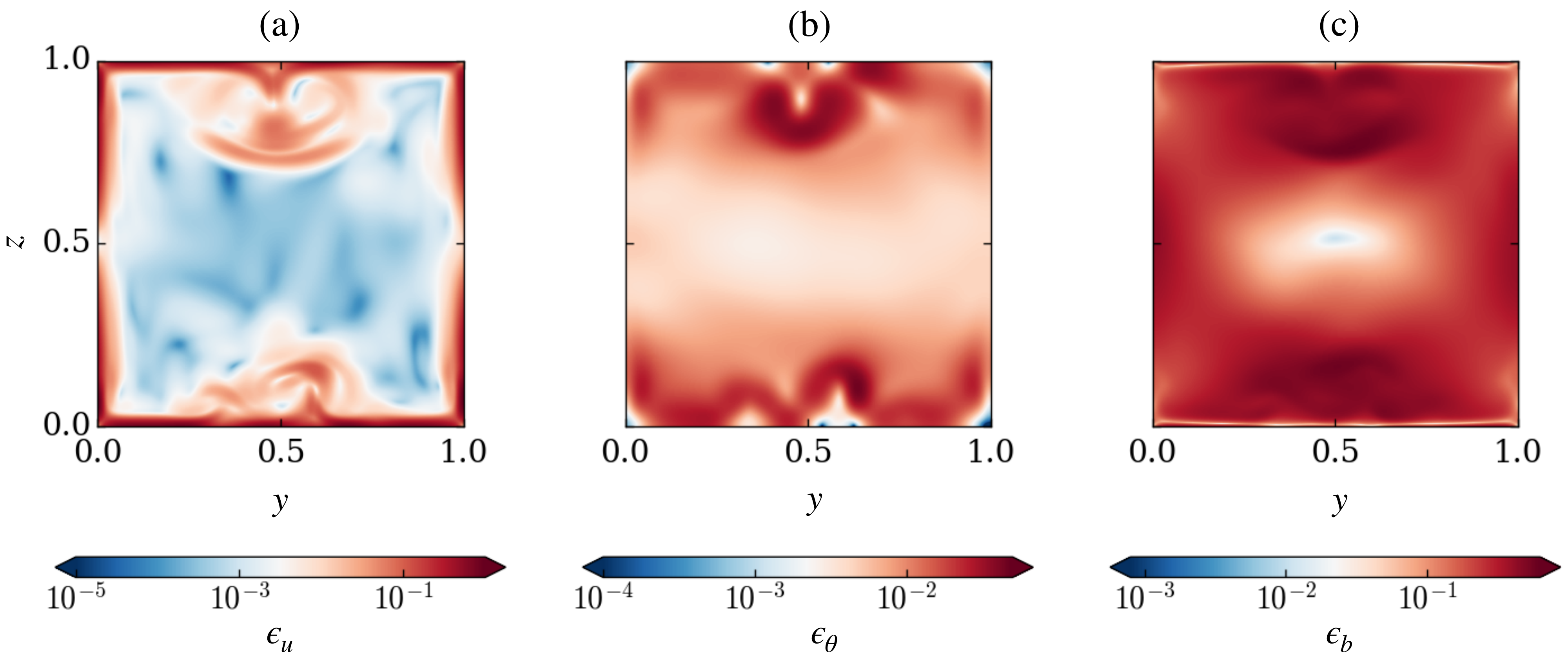}
    \caption{For $\mathrm{Ra}=10^5$, $\mathrm{Ha}=20$ (Case II), Contours of (a) viscous dissipation rate, (b) thermal dissipation rate based on temperature fluctuations, and (c) magnetic dissipation rate on the vertical midplane ($x=0.5$). The viscous and thermal dissipation rates are more intense near the boundaries.}
    \label{fig:Dissipation}
\end{figure}

Thus far, we have examined the relative contributions of the bulk and boundary layers to the total dissipation rates. We now turn to their spatial distribution. An important geometric consideration is that the boundary layers occupy a small fraction of the total volume: the viscous boundary layers account for roughly 10\% of the volume, while the thermal boundary layers, being relatively thicker at low Prandtl numbers, account for roughly 30\%. Consequently, even though the total viscous and thermal dissipation rates (the latter based on temperature fluctuations) are of the same order in Cases~I and II, the dissipation is considerably more intense within the boundary layers.
This is evident in Figs.~\ref{fig:Dissipation}(a) and (b), which show contour plots of the local viscous and thermal dissipation rates, respectively, for Case II. The local viscous dissipation rate in the boundary layers is four to five orders of magnitude larger than in the bulk. The magnetic dissipation rate, by contrast, appears relatively uniform, with only small localized regions of reduced dissipation near the cell center and immediately adjacent to the walls.

We quantify this intensity by computing volume-averaged dissipation rates --- defined as the total dissipation in each region divided by its volume. The volume-averaged viscous and thermal dissipation rates in the boundary layers are several times larger than in the bulk, whereas the volume-averaged magnetic dissipation rates are of the same order in both regions for Cases~II and III, indicating comparable dissipation intensity throughout the domain.
\begin{figure}
    \centering
    \includegraphics[scale=0.25]{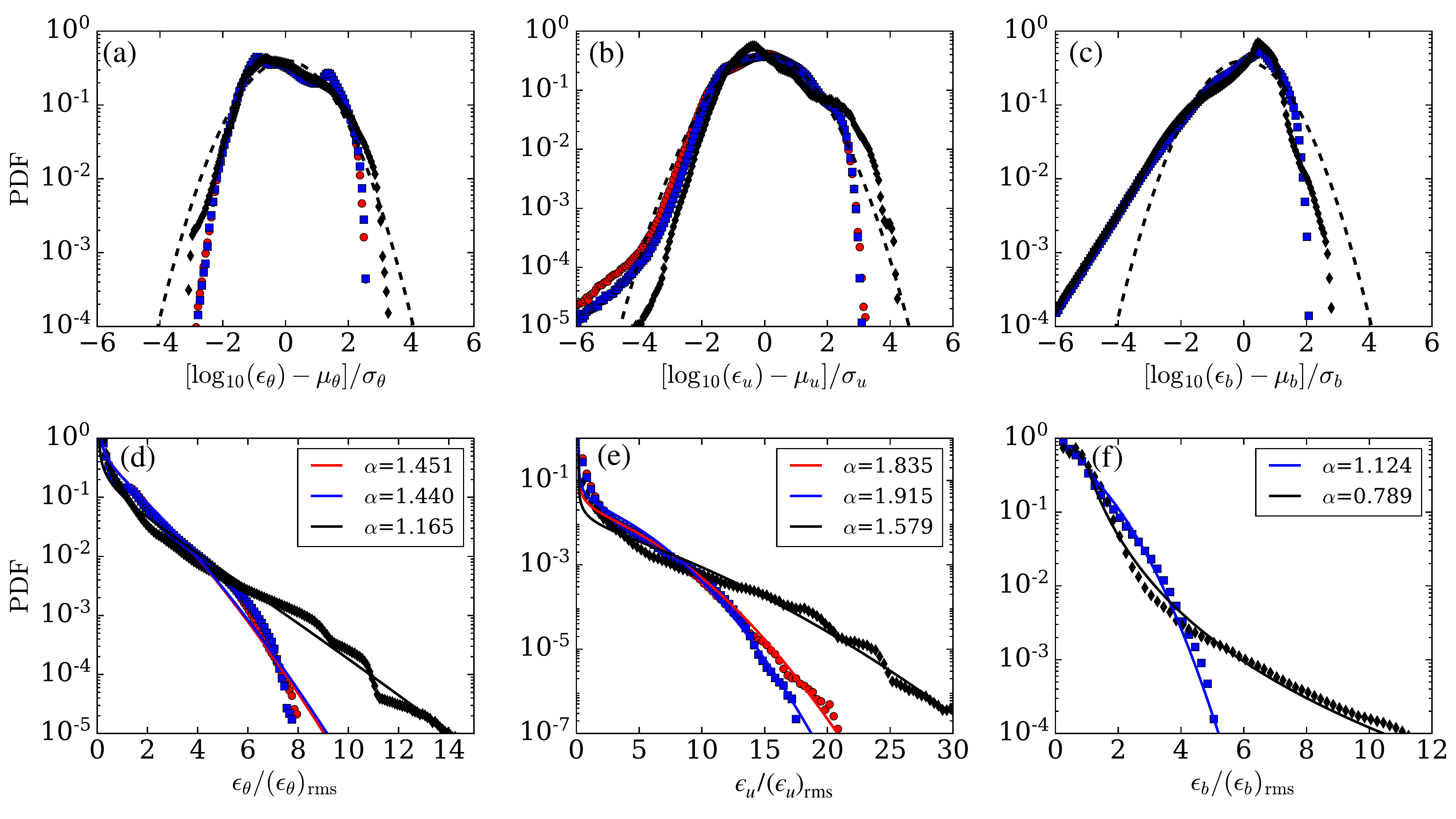}
    \caption{Probability density functions (PDFs) of the dissipation rates. Top panel: PDFs of $[\log_{10}(\epsilon_i) - \mu_i]/\sigma_i$ corresponding to (a) thermal, (b) viscous, and (c) magnetic dissipation rates for Case I (red circles), Case II (blue squares), and Case III (black diamonds). The dissipation rates deviate from log-normal profiles (dashed black curves). Bottom panel: PDFs of (d) thermal, (e) viscous, and (f) magnetic dissipation rates, normalized by their respective mean values, for Case I (red circles), Case II (blue squares), and Case III (black diamonds). The PDFs fit well with the stretched exponential curve, with the stretching exponents $\alpha$ decreasing with increasing Hartmann number.}
    \label{fig:PDF_Dissipation}
\end{figure}

We further characterize the dissipation rates by computing the probability density functions (PDFs) of the local viscous, thermal, and magnetic dissipation rates. According to the refined similarity hypothesis of \citet{Kolmogorov:JFM1963}, dissipation fields in turbulent flows are expected to follow a log-normal distribution. Figures~\ref{fig:PDF_Dissipation}(a), (b), and (c) show the PDFs of the thermal (based on $\theta$), viscous, and magnetic dissipation rates, respectively, plotted in log-normal coordinates, where $\mu_i$ and $\sigma_i$ denote the mean and standard deviation of the corresponding dissipation rate. All three dissipation rates exhibit clear deviations from the log-normal profile (shown by the black dashed curves), most prominently for the magnetic dissipation rate. Such deviations are attributed to the intermittent nature of the local viscous and thermal dissipation fields and have been reported in earlier in the studies of passive scalar turbulence~\cite{Schumacher:PF2005} and Rayleigh-Bénard convection \cite{Emran:JFM2008,Kaczorowski:JFM2008,Zhang:JFM2017}.

Figures~\ref{fig:PDF_Dissipation}(d), (e), and (f) show the PDFs of the thermal, viscous, and magnetic dissipation rates, respectively, each normalized by their mean value. The PDFs exhibit extended tails that become more pronounced with increasing Hartmann number. Consistent with earlier studies of viscous and thermal dissipation rates in Rayleigh--B\'enard convection~\cite{Emran:JFM2008,Zhang:JFM2017,Bhattacharya:PF2018,Bhattacharya:PF2019}, the PDFs are well described by the stretched exponential function
\begin{equation}
    p(Y) = \frac{C}{\sqrt{Y}} \exp(-m Y^\alpha),
\end{equation}
where $m = 0.38$, $0.042$, and $1.3$ for the thermal, viscous, and magnetic dissipation rates, respectively, and $\alpha$ decreases with increasing Hartmann number for each dissipation rate. The extended tails in the viscous dissipation PDFs are associated with the boundary layers, where viscous dissipation is most intense, while those in the thermal dissipation PDFs (based on $\theta$) reflect stronger temperature fluctuations in the bulk. The growth of these tails with increasing Hartmann number is consistent with the increasing boundary-layer dominance of the viscous and magnetic dissipation rates and the bulk dominance of the thermal dissipation rates as reported in Table~\ref{table:Dissipation}.

We conclude this section by discussing why magnetic fields promote extreme dissipation events. Even when the Reynolds number is held fixed by appropriately adjusting the Rayleigh number, the magnetic field further thins the viscous boundary layer, steepening the velocity gradients within it and thereby increasing the local viscous dissipation rate. The magnetic dissipation, being proportional to the square of the induced currents --- which scale linearly with the Hartmann number~\cite{Chandrasekhar:book:Instability,Davidson:book:MHD} --- is further amplified by the confinement of these currents within the thin boundary layers, where they must close to satisfy the divergence-free condition. Consequently, both the viscous and magnetic dissipation rates in the boundary layers intensify with increasing Hartmann number. It should be noted, however, that despite this intensification, the volume-averaged magnetic dissipation intensity remains of the same order in the boundary layers and the bulk.

We conclude in the next section.

\section{Summary and Conclusions} \label{sec:Conclusion}
In this paper, we employed direct numerical simulations to investigate the impact of a vertical magnetic field on the flow structures, heat transport, and dissipation rates in thermal convection, holding the Reynolds number approximately fixed across three cases with Hartmann numbers $\mathrm{Ha} = 0$, $20$, and $220$. The Reynolds number, which quantifies the intensity of turbulent fluctuations, was maintained at $\mathrm{Re} \approx 880$ in all three cases by appropriately adjusting the Rayleigh number.
Despite comparable levels of turbulence intensity, the differing magnetic field strengths give rise to pronounced differences in the flow structures and the resulting heat transport and dissipation rates, as summarised below.

The large-scale circulation roll becomes elongated along the vertical direction with increasing magnetic field strength and eventually disintegrates into multiple elongated convection rolls, rendering the flow increasingly anisotropic. Concurrently, the viscous boundary layer thins with increasing magnetic field strength due to the formation of Hartmann and Shercliff layers.

The Nusselt number increases modestly with increasing magnetic field strength, indicating that the thermal boundary layers are also thinned by the magnetic field. The PDF of the convective heat flux develops increasingly pronounced right tails at higher field strengths, reflecting an amplification of positive heat flux events that drives the higher Nusselt number. Thus, although strong magnetic fields suppress turbulent mixing, they enhance heat transport by amplifying the relative contribution of extreme positive heat flux events to the global heat flux.

For weak or absent magnetic fields, the viscous dissipation rates in the bulk and boundary layers are of the same order, while the thermal dissipation is strongly boundary-layer dominated due to steep near-wall temperature gradients. For strong magnetic fields ($\mathrm{Ha} = 220$), the boundary-layer contribution to the viscous dissipation increases further, whereas the thermal boundary-layer contribution weakens considerably. Decomposing the thermal dissipation into contributions from the mean temperature profile ($\tilde{D}_{\bar{T}}$) and temperature fluctuations ($\tilde{D}_\theta$) revealed that the reduction in boundary-layer thermal dissipation at large Hartmann numbers is driven by a suppression of temperature fluctuations in the boundary layer relative to the bulk. The magnetic dissipation, by contrast, is strongly bulk-dominated in all magnetoconvection cases, as the electrically insulating walls suppress near-wall currents.

We further examine the spatial distribution and statistics of the dissipation rates. Since the boundary layers occupy only a small fraction of the total volume (roughly 10\% for viscous and 30\% for thermal boundary layers), the volume-averaged viscous and thermal dissipation rates are considerably more intense in the boundary layers than in the bulk, whereas the volume-averaged magnetic dissipation intensity is comparable across both regions. The PDFs of all three dissipation rates deviate from the log-normal profile and are well described by a stretched exponential function, with increasingly pronounced tails at higher Hartmann numbers. The growth of these tails reflects the intensification of viscous and magnetic dissipation in the boundary layers with increasing field strength, driven by boundary-layer thinning and the confinement of induced currents. Despite this intensification, the volume-averaged magnetic dissipation intensity remains of the same order in both regions.

Our study provides insights into the Hartmann number dependence of flow structures, boundary layers, heat transport, and dissipation rates in turbulent magnetoconvection at fixed Reynolds number. These results are relevant to a range of technological applications in which magnetic fields are used to control turbulent flows with heat transfer, including liquid metal blanket cooling in fusion reactors and magnetic braking in continuous casting processes. The present analysis is restricted to convection under a vertical magnetic field; extending this to horizontal magnetic field configurations is a natural direction for future work.

\begin{acknowledgments}
The authors thank G. Bhutani, P. Kumar, and A. S. Dwivedi for useful discussions. AB and SB acknowledge Anusandhan National Research Foundation for financial support under the grant number ANRF/ECRG/2024/000809/ENS. SB thanks IIT Mandi for the Seed Grant number IITM/SG/SBh/137.
\end{acknowledgments}

\section*{data availability}
The data that support the findings of this study are available on reasonable request.

%\bibliography{bib/covid19,bib/journal,bib/book,bib/book_chapter,bib/web,bib/preprint,bib/report}

\begin{thebibliography}{57}%
	\makeatletter
	\providecommand \@ifxundefined [1]{%
		\@ifx{#1\undefined}
	}%
	\providecommand \@ifnum [1]{%
		\ifnum #1\expandafter \@firstoftwo
		\else \expandafter \@secondoftwo
		\fi
	}%
	\providecommand \@ifx [1]{%
		\ifx #1\expandafter \@firstoftwo
		\else \expandafter \@secondoftwo
		\fi
	}%
	\providecommand \natexlab [1]{#1}%
	\providecommand \enquote  [1]{``#1''}%
	\providecommand \bibnamefont  [1]{#1}%
	\providecommand \bibfnamefont [1]{#1}%
	\providecommand \citenamefont [1]{#1}%
	\providecommand \href@noop [0]{\@secondoftwo}%
	\providecommand \href [0]{\begingroup \@sanitize@url \@href}%
	\providecommand \@href[1]{\@@startlink{#1}\@@href}%
	\providecommand \@@href[1]{\endgroup#1\@@endlink}%
	\providecommand \@sanitize@url [0]{\catcode `\\12\catcode `\$12\catcode
		`\&12\catcode `\#12\catcode `\^12\catcode `\_12\catcode `\%12\relax}%
	\providecommand \@@startlink[1]{}%
	\providecommand \@@endlink[0]{}%
	\providecommand \url  [0]{\begingroup\@sanitize@url \@url }%
	\providecommand \@url [1]{\endgroup\@href {#1}{\urlprefix }}%
	\providecommand \urlprefix  [0]{URL }%
	\providecommand \Eprint [0]{\href }%
	\providecommand \doibase [0]{https://doi.org/}%
	\providecommand \selectlanguage [0]{\@gobble}%
	\providecommand \bibinfo  [0]{\@secondoftwo}%
	\providecommand \bibfield  [0]{\@secondoftwo}%
	\providecommand \translation [1]{[#1]}%
	\providecommand \BibitemOpen [0]{}%
	\providecommand \bibitemStop [0]{}%
	\providecommand \bibitemNoStop [0]{.\EOS\space}%
	\providecommand \EOS [0]{\spacefactor3000\relax}%
	\providecommand \BibitemShut  [1]{\csname bibitem#1\endcsname}%
	\let\auto@bib@innerbib\@empty
	%</preamble>
	\bibitem [{\citenamefont
		{Chandrasekhar}(1981)}]{Chandrasekhar:book:Instability}%
	\BibitemOpen
	\bibfield  {author} {\bibinfo {author} {\bibfnamefont {S.}~\bibnamefont
			{Chandrasekhar}},\ }\href@noop {} {\emph {\bibinfo {title} {{Hydrodynamic and
					Hydromagnetic Stability}}}}\ (\bibinfo  {publisher} {Dover publications},\
	\bibinfo {address} {Oxford},\ \bibinfo {year} {1981})\BibitemShut {NoStop}%
	\bibitem [{\citenamefont {Weiss}\ and\ \citenamefont
		{Proctor}(2014)}]{Weiss:book}%
	\BibitemOpen
	\bibfield  {author} {\bibinfo {author} {\bibfnamefont {N.~O.}\ \bibnamefont
			{Weiss}}\ and\ \bibinfo {author} {\bibfnamefont {M.~R.~E.}\ \bibnamefont
			{Proctor}},\ }\href@noop {} {\emph {\bibinfo {title} {{Magnetoconvection}}}}\
	(\bibinfo  {publisher} {Cambridge University Press},\ \bibinfo {address}
	{Cambridge},\ \bibinfo {year} {2014})\BibitemShut {NoStop}%
	\bibitem [{\citenamefont {Davidson}(2017)}]{Davidson:book:MHD}%
	\BibitemOpen
	\bibfield  {author} {\bibinfo {author} {\bibfnamefont {P.~A.}\ \bibnamefont
			{Davidson}},\ }\href@noop {} {\emph {\bibinfo {title} {{An introduction to
					magnetohydrodynamics}}}},\ \bibinfo {edition} {2nd}\ ed.\ (\bibinfo
	{publisher} {Cambridge University Press},\ \bibinfo {address} {Cambridge},\
	\bibinfo {year} {2017})\BibitemShut {NoStop}%
	\bibitem [{\citenamefont {Schumacher}\ and\ \citenamefont
		{Sreenivasan}(2020)}]{Schumacher:RMP2020}%
	\BibitemOpen
	\bibfield  {author} {\bibinfo {author} {\bibfnamefont {J.}~\bibnamefont
			{Schumacher}}\ and\ \bibinfo {author} {\bibfnamefont {K.~R.}\ \bibnamefont
			{Sreenivasan}},\ }\bibfield  {title} {\bibinfo {title} {{Colloquium: Unusual
				dynamics of convection in the Sun}},\ }\href@noop {} {\bibfield  {journal}
		{\bibinfo  {journal} {Rev. Mod. Phys.}\ }\textbf {\bibinfo {volume} {92}},\
		\bibinfo {pages} {041001} (\bibinfo {year} {2020})}\BibitemShut {NoStop}%
	\bibitem [{\citenamefont {Stefani}(2024)}]{Stefani:NRP2024}%
	\BibitemOpen
	\bibfield  {author} {\bibinfo {author} {\bibfnamefont {F.}~\bibnamefont
			{Stefani}},\ }\bibfield  {title} {\bibinfo {title} {{Liquid-metal experiments
				on geophysical and astrophysical phenomena}},\ }\href@noop {} {\bibfield
		{journal} {\bibinfo  {journal} {Nat. Rev. Phys.}\ }\textbf {\bibinfo {volume}
			{6}},\ \bibinfo {pages} {409} (\bibinfo {year} {2024})}\BibitemShut {NoStop}%
	\bibitem [{\citenamefont {Davidson}(1999)}]{Davidson:ARFM1999}%
	\BibitemOpen
	\bibfield  {author} {\bibinfo {author} {\bibfnamefont {P.~A.}\ \bibnamefont
			{Davidson}},\ }\bibfield  {title} {\bibinfo {title} {{Magnetohydrodynamics in
				materials processing}},\ }\href@noop {} {\bibfield  {journal} {\bibinfo
			{journal} {Annu. Rev. Fluid Mech.}\ }\textbf {\bibinfo {volume} {31}},\
		\bibinfo {pages} {273} (\bibinfo {year} {1999})}\BibitemShut {NoStop}%
	\bibitem [{\citenamefont {Lyubimov}\ \emph {et~al.}(2010)\citenamefont
		{Lyubimov}, \citenamefont {Burnysheva}, \citenamefont {Benhadid},
		\citenamefont {Lyubimova},\ and\ \citenamefont {Henry}}]{Lyubimov:JFM2010}%
	\BibitemOpen
	\bibfield  {author} {\bibinfo {author} {\bibfnamefont {D.~V.}\ \bibnamefont
			{Lyubimov}}, \bibinfo {author} {\bibfnamefont {A.~V.}\ \bibnamefont
			{Burnysheva}}, \bibinfo {author} {\bibfnamefont {H.}~\bibnamefont
			{Benhadid}}, \bibinfo {author} {\bibfnamefont {T.~P.}\ \bibnamefont
			{Lyubimova}},\ and\ \bibinfo {author} {\bibfnamefont {D.}~\bibnamefont
			{Henry}},\ }\bibfield  {title} {\bibinfo {title} {{Rotating magnetic field
				effect on convection and its stability in a horizontal cylinder subjected to
				a longitudinal temperature gradient}},\ }\href@noop {} {\bibfield  {journal}
		{\bibinfo  {journal} {J. Fluid Mech.}\ }\textbf {\bibinfo {volume} {664}},\
		\bibinfo {pages} {108} (\bibinfo {year} {2010})}\BibitemShut {NoStop}%
	\bibitem [{\citenamefont {Kelley}\ and\ \citenamefont
		{Sadoway}(2014)}]{Kelley:PF2014}%
	\BibitemOpen
	\bibfield  {author} {\bibinfo {author} {\bibfnamefont {D.}~\bibnamefont
			{Kelley}}\ and\ \bibinfo {author} {\bibfnamefont {D.~R.}\ \bibnamefont
			{Sadoway}},\ }\bibfield  {title} {\bibinfo {title} {{Mixing in a liquid metal
				electrode}},\ }\href@noop {} {\bibfield  {journal} {\bibinfo  {journal}
			{Phys. Fluids}\ }\textbf {\bibinfo {volume} {26}},\ \bibinfo {pages} {057102}
		(\bibinfo {year} {2014})}\BibitemShut {NoStop}%
	\bibitem [{\citenamefont {Shen}\ and\ \citenamefont
		{Zikanov}(2016)}]{Shen:TCFD2016}%
	\BibitemOpen
	\bibfield  {author} {\bibinfo {author} {\bibfnamefont {Y.}~\bibnamefont
			{Shen}}\ and\ \bibinfo {author} {\bibfnamefont {O.}~\bibnamefont {Zikanov}},\
	}\bibfield  {title} {\bibinfo {title} {{Thermal convection in a liquid metal
				battery}},\ }\href@noop {} {\bibfield  {journal} {\bibinfo  {journal} {Theor.
				Comput. Fluid Dyn.}\ }\textbf {\bibinfo {volume} {30}},\ \bibinfo {pages}
		{275} (\bibinfo {year} {2016})}\BibitemShut {NoStop}%
	\bibitem [{\citenamefont {Kelley}\ and\ \citenamefont
		{Weier}(2018)}]{Kelley:AMR2018}%
	\BibitemOpen
	\bibfield  {author} {\bibinfo {author} {\bibfnamefont {D.}~\bibnamefont
			{Kelley}}\ and\ \bibinfo {author} {\bibfnamefont {T.}~\bibnamefont {Weier}},\
	}\bibfield  {title} {\bibinfo {title} {{Fluid Mechanics of Liquid Metal
				Batteries}},\ }\href@noop {} {\bibfield  {journal} {\bibinfo  {journal}
			{Appl. Mech. Rev.}\ }\textbf {\bibinfo {volume} {70}},\ \bibinfo {pages}
		{020801} (\bibinfo {year} {2018})}\BibitemShut {NoStop}%
	\bibitem [{\citenamefont {Davidson}\ \emph {et~al.}(2022)\citenamefont
		{Davidson}, \citenamefont {Wong}, \citenamefont {Atkinson},\ and\
		\citenamefont {Ranjan}}]{Davidson:PRF2022}%
	\BibitemOpen
	\bibfield  {author} {\bibinfo {author} {\bibfnamefont {P.~A.}\ \bibnamefont
			{Davidson}}, \bibinfo {author} {\bibfnamefont {O.}~\bibnamefont {Wong}},
		\bibinfo {author} {\bibfnamefont {J.~W.}\ \bibnamefont {Atkinson}},\ and\
		\bibinfo {author} {\bibfnamefont {A.}~\bibnamefont {Ranjan}},\ }\bibfield
	{title} {\bibinfo {title} {{Magnetically driven flow in a liquid-metal
				battery}},\ }\href@noop {} {\bibfield  {journal} {\bibinfo  {journal} {Phys.
				Rev. Fluids}\ }\textbf {\bibinfo {volume} {7}},\ \bibinfo {pages} {074701}
		(\bibinfo {year} {2022})}\BibitemShut {NoStop}%
	\bibitem [{\citenamefont {Buhler}\ and\ \citenamefont
		{Mistrangelo}(2013)}]{Buhler:FED2013}%
	\BibitemOpen
	\bibfield  {author} {\bibinfo {author} {\bibfnamefont {L.}~\bibnamefont
			{Buhler}}\ and\ \bibinfo {author} {\bibfnamefont {C.}~\bibnamefont
			{Mistrangelo}},\ }\bibfield  {title} {\bibinfo {title} {{Magnetohydrodynamic
				flows in breeder units of a HCLL blanket with spatially varying magnetic
				fields}},\ }\href@noop {} {\bibfield  {journal} {\bibinfo  {journal} {Fusion
				Eng. Des.}\ }\textbf {\bibinfo {volume} {88}},\ \bibinfo {pages} {2314}
		(\bibinfo {year} {2013})}\BibitemShut {NoStop}%
	\bibitem [{\citenamefont {Mistrangelo}\ \emph {et~al.}(2020)\citenamefont
		{Mistrangelo}, \citenamefont {B{\"u}hler},\ and\ \citenamefont
		{Kl{\"u}ber}}]{Mistrangelo:FED2020}%
	\BibitemOpen
	\bibfield  {author} {\bibinfo {author} {\bibfnamefont {C.}~\bibnamefont
			{Mistrangelo}}, \bibinfo {author} {\bibfnamefont {L.}~\bibnamefont
			{B{\"u}hler}},\ and\ \bibinfo {author} {\bibfnamefont {V.}~\bibnamefont
			{Kl{\"u}ber}},\ }\bibfield  {title} {\bibinfo {title} {{Three-dimensional
				magneto convective flows in geometries relevant for DCLL blankets}},\
	}\href@noop {} {\bibfield  {journal} {\bibinfo  {journal} {Fusion Eng. Des.}\
		}\textbf {\bibinfo {volume} {159}},\ \bibinfo {pages} {111686} (\bibinfo
		{year} {2020})}\BibitemShut {NoStop}%
	\bibitem [{\citenamefont {Mistrangelo}\ \emph {et~al.}(2021)\citenamefont
		{Mistrangelo}, \citenamefont {B{\"u}hler}, \citenamefont {Smolentsev},
		\citenamefont {Kl{\"u}ber}, \citenamefont {Maione},\ and\ \citenamefont
		{Aubert}}]{Mistrangelo:FED2021}%
	\BibitemOpen
	\bibfield  {author} {\bibinfo {author} {\bibfnamefont {C.}~\bibnamefont
			{Mistrangelo}}, \bibinfo {author} {\bibfnamefont {L.}~\bibnamefont
			{B{\"u}hler}}, \bibinfo {author} {\bibfnamefont {S.}~\bibnamefont
			{Smolentsev}}, \bibinfo {author} {\bibfnamefont {V.}~\bibnamefont
			{Kl{\"u}ber}}, \bibinfo {author} {\bibfnamefont {I.}~\bibnamefont {Maione}},\
		and\ \bibinfo {author} {\bibfnamefont {J.}~\bibnamefont {Aubert}},\
	}\bibfield  {title} {\bibinfo {title} {{MHD flow in liquid metal blankets:
				Major design issues, MHD guidelines and numerical analysis}},\ }\href@noop {}
	{\bibfield  {journal} {\bibinfo  {journal} {Fusion Eng. Des.}\ }\textbf
		{\bibinfo {volume} {173}},\ \bibinfo {pages} {112795} (\bibinfo {year}
		{2021})}\BibitemShut {NoStop}%
	\bibitem [{\citenamefont {Roberts}(1967)}]{Roberts:book}%
	\BibitemOpen
	\bibfield  {author} {\bibinfo {author} {\bibfnamefont {P.~H.}\ \bibnamefont
			{Roberts}},\ }\href@noop {} {\emph {\bibinfo {title} {{An introduction to
					magnetohydrodynamics}}}}\ (\bibinfo  {publisher} {Longmans},\ \bibinfo
	{address} {London},\ \bibinfo {year} {1967})\BibitemShut {NoStop}%
	\bibitem [{\citenamefont {Verma}(2019)}]{Verma:book:ET}%
	\BibitemOpen
	\bibfield  {author} {\bibinfo {author} {\bibfnamefont {M.~K.}\ \bibnamefont
			{Verma}},\ }\href@noop {} {\emph {\bibinfo {title} {Energy trasnfers in Fluid
				Flows: Multiscale and Spectral Perspectives}}}\ (\bibinfo  {publisher}
	{Cambridge University Press},\ \bibinfo {address} {Cambridge},\ \bibinfo
	{year} {2019})\BibitemShut {NoStop}%
	\bibitem [{\citenamefont {Fauve}\ \emph {et~al.}(1981)\citenamefont {Fauve},
		\citenamefont {Laroche},\ and\ \citenamefont {Libchaber}}]{Fauve:JPL1981}%
	\BibitemOpen
	\bibfield  {author} {\bibinfo {author} {\bibfnamefont {S.}~\bibnamefont
			{Fauve}}, \bibinfo {author} {\bibfnamefont {C.}~\bibnamefont {Laroche}},\
		and\ \bibinfo {author} {\bibfnamefont {A.}~\bibnamefont {Libchaber}},\
	}\bibfield  {title} {\bibinfo {title} {{Effect of a horizontal magnetic field
				on convective instabilities in mercury}},\ }\href@noop {} {\bibfield
		{journal} {\bibinfo  {journal} {J. Physique Lett.}\ }\textbf {\bibinfo
			{volume} {42}},\ \bibinfo {pages} {L455} (\bibinfo {year}
		{1981})}\BibitemShut {NoStop}%
	\bibitem [{\citenamefont {Busse}\ and\ \citenamefont
		{Clever}(1983)}]{Busse:JTAM1983}%
	\BibitemOpen
	\bibfield  {author} {\bibinfo {author} {\bibfnamefont {F.~H.}\ \bibnamefont
			{Busse}}\ and\ \bibinfo {author} {\bibfnamefont {R.~M.}\ \bibnamefont
			{Clever}},\ }\bibfield  {title} {\bibinfo {title} {{Stability of convection
				rolls in the presence of a horizontal magnetic field}},\ }\href@noop {}
	{\bibfield  {journal} {\bibinfo  {journal} {J. Theor. Appl. Mech.}\ }\textbf
		{\bibinfo {volume} {2}},\ \bibinfo {pages} {495} (\bibinfo {year}
		{1983})}\BibitemShut {NoStop}%
	\bibitem [{\citenamefont {Vogt}\ \emph {et~al.}(2018)\citenamefont {Vogt},
		\citenamefont {Ishimi}, \citenamefont {Yanagisawa}, \citenamefont {Tasaka},
		\citenamefont {Sakuraba},\ and\ \citenamefont {Eckert}}]{Vogt:PRF2018}%
	\BibitemOpen
	\bibfield  {author} {\bibinfo {author} {\bibfnamefont {T.}~\bibnamefont
			{Vogt}}, \bibinfo {author} {\bibfnamefont {W.}~\bibnamefont {Ishimi}},
		\bibinfo {author} {\bibfnamefont {T.}~\bibnamefont {Yanagisawa}}, \bibinfo
		{author} {\bibfnamefont {Y.}~\bibnamefont {Tasaka}}, \bibinfo {author}
		{\bibfnamefont {A.}~\bibnamefont {Sakuraba}},\ and\ \bibinfo {author}
		{\bibfnamefont {S.}~\bibnamefont {Eckert}},\ }\bibfield  {title} {\bibinfo
		{title} {{Transition between quasi-two-dimensional and three-dimensional
				Rayleigh-B{\'e}nard convection in a horizontal magnetic field}},\ }\href@noop
	{} {\bibfield  {journal} {\bibinfo  {journal} {Phys. Rev. Fluids}\ }\textbf
		{\bibinfo {volume} {3}},\ \bibinfo {pages} {013503} (\bibinfo {year}
		{2018})}\BibitemShut {NoStop}%
	\bibitem [{\citenamefont {Cioni}\ \emph {et~al.}(1997)\citenamefont {Cioni},
		\citenamefont {Ciliberto},\ and\ \citenamefont {Sommeria}}]{Cioni:JFM1997}%
	\BibitemOpen
	\bibfield  {author} {\bibinfo {author} {\bibfnamefont {S.}~\bibnamefont
			{Cioni}}, \bibinfo {author} {\bibfnamefont {S.}~\bibnamefont {Ciliberto}},\
		and\ \bibinfo {author} {\bibfnamefont {J.}~\bibnamefont {Sommeria}},\
	}\bibfield  {title} {\bibinfo {title} {{Strongly turbulent
				Rayleigh{\textendash}B{\'e}nard convection in mercury: comparison with
				results at moderate Prandtl number}},\ }\href@noop {} {\bibfield  {journal}
		{\bibinfo  {journal} {J. Fluid Mech.}\ }\textbf {\bibinfo {volume} {335}},\
		\bibinfo {pages} {111} (\bibinfo {year} {1997})}\BibitemShut {NoStop}%
	\bibitem [{\citenamefont {King}\ and\ \citenamefont
		{Aurnou}(2015)}]{King:PNAS2015}%
	\BibitemOpen
	\bibfield  {author} {\bibinfo {author} {\bibfnamefont {E.~M.}\ \bibnamefont
			{King}}\ and\ \bibinfo {author} {\bibfnamefont {J.~M.}\ \bibnamefont
			{Aurnou}},\ }\bibfield  {title} {\bibinfo {title} {{Magnetostrophic balance
				as the optimal state for turbulent magnetoconvection}},\ }\href@noop {}
	{\bibfield  {journal} {\bibinfo  {journal} {Proc. Natl. Acad. Sci. USA}\
		}\textbf {\bibinfo {volume} {112}},\ \bibinfo {pages} {990} (\bibinfo {year}
		{2015})}\BibitemShut {NoStop}%
	\bibitem [{\citenamefont {Busse}(2008)}]{Busse:PF2008}%
	\BibitemOpen
	\bibfield  {author} {\bibinfo {author} {\bibfnamefont {F.~H.}\ \bibnamefont
			{Busse}},\ }\bibfield  {title} {\bibinfo {title} {{Asymptotic theory of
				wall-attached convection in a horizontal fluid layer with a vertical magnetic
				field}},\ }\href@noop {} {\bibfield  {journal} {\bibinfo  {journal} {Phys.
				Fluids}\ }\textbf {\bibinfo {volume} {20}},\ \bibinfo {pages} {024102}
		(\bibinfo {year} {2008})}\BibitemShut {NoStop}%
	\bibitem [{\citenamefont {Liu}\ \emph {et~al.}(2018)\citenamefont {Liu},
		\citenamefont {Krasnov},\ and\ \citenamefont {Schumacher}}]{Liu:JFM2018}%
	\BibitemOpen
	\bibfield  {author} {\bibinfo {author} {\bibfnamefont {W.}~\bibnamefont
			{Liu}}, \bibinfo {author} {\bibfnamefont {D.}~\bibnamefont {Krasnov}},\ and\
		\bibinfo {author} {\bibfnamefont {J.}~\bibnamefont {Schumacher}},\ }\bibfield
	{title} {\bibinfo {title} {{Wall modes in magnetoconvection at high Hartmann
				numbers}},\ }\href@noop {} {\bibfield  {journal} {\bibinfo  {journal} {J.
				Fluid Mech.}\ }\textbf {\bibinfo {volume} {849}},\ \bibinfo {pages} {R2}
		(\bibinfo {year} {2018})}\BibitemShut {NoStop}%
	\bibitem [{\citenamefont {Z\"urner}\ \emph {et~al.}(2020)\citenamefont
		{Z\"urner}, \citenamefont {Schindler}, \citenamefont {Vogt}, \citenamefont
		{Eckert},\ and\ \citenamefont {J\"org}}]{Zuerner:JFM2020}%
	\BibitemOpen
	\bibfield  {author} {\bibinfo {author} {\bibfnamefont {T.}~\bibnamefont
			{Z\"urner}}, \bibinfo {author} {\bibfnamefont {F.}~\bibnamefont {Schindler}},
		\bibinfo {author} {\bibfnamefont {T.}~\bibnamefont {Vogt}}, \bibinfo {author}
		{\bibfnamefont {S.}~\bibnamefont {Eckert}},\ and\ \bibinfo {author}
		{\bibfnamefont {S.}~\bibnamefont {J\"org}},\ }\bibfield  {title} {\bibinfo
		{title} {{Flow regimes of Rayleigh-B{\'e}nard convection in a vertical
				magnetic field}},\ }\href@noop {} {\bibfield  {journal} {\bibinfo  {journal}
			{J. Fluid Mech.}\ }\textbf {\bibinfo {volume} {894}},\ \bibinfo {pages} {A21}
		(\bibinfo {year} {2020})}\BibitemShut {NoStop}%
	\bibitem [{\citenamefont {Akhmedagaev}\ \emph {et~al.}(2020)\citenamefont
		{Akhmedagaev}, \citenamefont {Zikanov}, \citenamefont {Krasnov},\ and\
		\citenamefont {Schumacher}}]{Akhmedagaev:JFM2020}%
	\BibitemOpen
	\bibfield  {author} {\bibinfo {author} {\bibfnamefont {R.}~\bibnamefont
			{Akhmedagaev}}, \bibinfo {author} {\bibfnamefont {O.}~\bibnamefont
			{Zikanov}}, \bibinfo {author} {\bibfnamefont {D.}~\bibnamefont {Krasnov}},\
		and\ \bibinfo {author} {\bibfnamefont {J.}~\bibnamefont {Schumacher}},\
	}\bibfield  {title} {\bibinfo {title} {{Turbulent Rayleigh-B{\'e}nard
				convection in a strong vertical magnetic field}},\ }\href@noop {} {\bibfield
		{journal} {\bibinfo  {journal} {J. Fluid Mech.}\ }\textbf {\bibinfo {volume}
			{895}},\ \bibinfo {pages} {R4} (\bibinfo {year} {2020})}\BibitemShut
	{NoStop}%
	\bibitem [{\citenamefont {Bhattacharya}\ \emph {et~al.}(2023)\citenamefont
		{Bhattacharya}, \citenamefont {Boeck}, \citenamefont {Krasnov},\ and\
		\citenamefont {Schumacher}}]{Bhattacharya:JFM2023}%
	\BibitemOpen
	\bibfield  {author} {\bibinfo {author} {\bibfnamefont {S.}~\bibnamefont
			{Bhattacharya}}, \bibinfo {author} {\bibfnamefont {T.}~\bibnamefont {Boeck}},
		\bibinfo {author} {\bibfnamefont {D.}~\bibnamefont {Krasnov}},\ and\ \bibinfo
		{author} {\bibfnamefont {J.}~\bibnamefont {Schumacher}},\ }\bibfield  {title}
	{\bibinfo {title} {{Effects of strong fringing magnetic fields on turbulent
				thermal convection}},\ }\href@noop {} {\bibfield  {journal} {\bibinfo
			{journal} {J. Fluid Mech.}\ }\textbf {\bibinfo {volume} {964}},\ \bibinfo
		{pages} {A31} (\bibinfo {year} {2023})}\BibitemShut {NoStop}%
	\bibitem [{\citenamefont {Bhattacharya}\ \emph {et~al.}(2024)\citenamefont
		{Bhattacharya}, \citenamefont {Boeck}, \citenamefont {Krasnov},\ and\
		\citenamefont {Schumacher}}]{Bhattacharya:JFM2024}%
	\BibitemOpen
	\bibfield  {author} {\bibinfo {author} {\bibfnamefont {S.}~\bibnamefont
			{Bhattacharya}}, \bibinfo {author} {\bibfnamefont {T.}~\bibnamefont {Boeck}},
		\bibinfo {author} {\bibfnamefont {D.}~\bibnamefont {Krasnov}},\ and\ \bibinfo
		{author} {\bibfnamefont {J.}~\bibnamefont {Schumacher}},\ }\bibfield  {title}
	{\bibinfo {title} {{Wall-attached convection under strong inclined magnetic
				fields}},\ }\href@noop {} {\bibfield  {journal} {\bibinfo  {journal} {J.
				Fluid Mech.}\ }\textbf {\bibinfo {volume} {979}},\ \bibinfo {pages} {A53}
		(\bibinfo {year} {2024})}\BibitemShut {NoStop}%
	\bibitem [{\citenamefont {McCormack}\ \emph {et~al.}(2025)\citenamefont
		{McCormack}, \citenamefont {Teimurazov}, \citenamefont {Shishkina},\ and\
		\citenamefont {Linkmann}}]{McCormack:JFM2025}%
	\BibitemOpen
	\bibfield  {author} {\bibinfo {author} {\bibfnamefont {M.}~\bibnamefont
			{McCormack}}, \bibinfo {author} {\bibfnamefont {A.}~\bibnamefont
			{Teimurazov}}, \bibinfo {author} {\bibfnamefont {O.}~\bibnamefont
			{Shishkina}},\ and\ \bibinfo {author} {\bibfnamefont {M.}~\bibnamefont
			{Linkmann}},\ }\bibfield  {title} {\bibinfo {title} {{Subcritical transition
				and multistability in liquid-metal magnetoconvection with sidewalls}},\
	}\href@noop {} {\bibfield  {journal} {\bibinfo  {journal} {J. Fluid Mech.}\
		}\textbf {\bibinfo {volume} {1023}},\ \bibinfo {pages} {R1} (\bibinfo {year}
		{2025})}\BibitemShut {NoStop}%
	\bibitem [{\citenamefont {Tao}\ \emph {et~al.}(2026)\citenamefont {Tao},
		\citenamefont {Zhu}, \citenamefont {Ni},\ and\ \citenamefont
		{Xie}}]{Tao:JFM2026}%
	\BibitemOpen
	\bibfield  {author} {\bibinfo {author} {\bibfnamefont {X.}~\bibnamefont
			{Tao}}, \bibinfo {author} {\bibfnamefont {X.}~\bibnamefont {Zhu}}, \bibinfo
		{author} {\bibfnamefont {M.-J.}\ \bibnamefont {Ni}},\ and\ \bibinfo {author}
		{\bibfnamefont {Y.-C.}\ \bibnamefont {Xie}},\ }\bibfield  {title} {\bibinfo
		{title} {{Onset and length scales of wall modes in confined magnetoconvection
				with a vertical magnetic field}},\ }\href@noop {} {\bibfield  {journal}
		{\bibinfo  {journal} {J. Fluid Mech.}\ }\textbf {\bibinfo {volume} {1026}},\
		\bibinfo {pages} {A41} (\bibinfo {year} {2026})}\BibitemShut {NoStop}%
	\bibitem [{\citenamefont {Hurlburt}\ \emph {et~al.}(1996)\citenamefont
		{Hurlburt}, \citenamefont {Matthews},\ and\ \citenamefont
		{Proctor}}]{Hurlburt:ApJ1996}%
	\BibitemOpen
	\bibfield  {author} {\bibinfo {author} {\bibfnamefont {N.}~\bibnamefont
			{Hurlburt}}, \bibinfo {author} {\bibfnamefont {P.}~\bibnamefont {Matthews}},\
		and\ \bibinfo {author} {\bibfnamefont {M.}~\bibnamefont {Proctor}},\
	}\bibfield  {title} {\bibinfo {title} {{Nonlinear compressible convection in
				oblique magnetic fields}},\ }\href@noop {} {\bibfield  {journal} {\bibinfo
			{journal} {Astrophys. J.}\ }\textbf {\bibinfo {volume} {457}},\ \bibinfo
		{pages} {933} (\bibinfo {year} {1996})}\BibitemShut {NoStop}%
	\bibitem [{\citenamefont {Nicoski}\ \emph {et~al.}(2022)\citenamefont
		{Nicoski}, \citenamefont {Yan},\ and\ \citenamefont
		{Calkins}}]{Nicoski:PRF2022}%
	\BibitemOpen
	\bibfield  {author} {\bibinfo {author} {\bibfnamefont {J.~A.}\ \bibnamefont
			{Nicoski}}, \bibinfo {author} {\bibfnamefont {M.}~\bibnamefont {Yan}},\ and\
		\bibinfo {author} {\bibfnamefont {M.~A.}\ \bibnamefont {Calkins}},\
	}\bibfield  {title} {\bibinfo {title} {{Quasistatic magnetoconvection with a
				tilted magnetic field}},\ }\href@noop {} {\bibfield  {journal} {\bibinfo
			{journal} {Phys. Rev. Fluids}\ }\textbf {\bibinfo {volume} {7}},\ \bibinfo
		{pages} {043504} (\bibinfo {year} {2022})}\BibitemShut {NoStop}%
	\bibitem [{\citenamefont {Yan}\ \emph {et~al.}(2019)\citenamefont {Yan},
		\citenamefont {Calkins}, \citenamefont {Maffei}, \citenamefont {Julien},
		\citenamefont {Tobias},\ and\ \citenamefont {Marti}}]{Yan:JFM2019}%
	\BibitemOpen
	\bibfield  {author} {\bibinfo {author} {\bibfnamefont {M.}~\bibnamefont
			{Yan}}, \bibinfo {author} {\bibfnamefont {M.}~\bibnamefont {Calkins}},
		\bibinfo {author} {\bibfnamefont {A.}~\bibnamefont {Maffei}}, \bibinfo
		{author} {\bibfnamefont {K.}~\bibnamefont {Julien}}, \bibinfo {author}
		{\bibfnamefont {S.}~\bibnamefont {Tobias}},\ and\ \bibinfo {author}
		{\bibfnamefont {P.}~\bibnamefont {Marti}},\ }\bibfield  {title} {\bibinfo
		{title} {{Heat transfer and flow regimes in quasi-static magnetoconvection
				with a vertical magnetic field}},\ }\href@noop {} {\bibfield  {journal}
		{\bibinfo  {journal} {J. Fluid Mech.}\ }\textbf {\bibinfo {volume} {877}},\
		\bibinfo {pages} {1186} (\bibinfo {year} {2019})}\BibitemShut {NoStop}%
	\bibitem [{\citenamefont {McCormack}\ \emph {et~al.}(2023)\citenamefont
		{McCormack}, \citenamefont {Teimurazov}, \citenamefont {Shishkina},\ and\
		\citenamefont {Linkmann}}]{McCormack:JFM2023}%
	\BibitemOpen
	\bibfield  {author} {\bibinfo {author} {\bibfnamefont {M.}~\bibnamefont
			{McCormack}}, \bibinfo {author} {\bibfnamefont {A.}~\bibnamefont
			{Teimurazov}}, \bibinfo {author} {\bibfnamefont {O.}~\bibnamefont
			{Shishkina}},\ and\ \bibinfo {author} {\bibfnamefont {M.}~\bibnamefont
			{Linkmann}},\ }\bibfield  {title} {\bibinfo {title} {{Wall mode dynamics and
				transition to chaos in magnetoconvection with a vertical magnetic field}},\
	}\href@noop {} {\bibfield  {journal} {\bibinfo  {journal} {J. Fluid Mech.}\
		}\textbf {\bibinfo {volume} {975}},\ \bibinfo {pages} {R2} (\bibinfo {year}
		{2023})}\BibitemShut {NoStop}%
	\bibitem [{\citenamefont {Teimurazov}\ \emph {et~al.}(2024)\citenamefont
		{Teimurazov}, \citenamefont {McCormack}, \citenamefont {Linkmann},\ and\
		\citenamefont {Shishkina}}]{Teimurazov:JFM2024}%
	\BibitemOpen
	\bibfield  {author} {\bibinfo {author} {\bibfnamefont {A.}~\bibnamefont
			{Teimurazov}}, \bibinfo {author} {\bibfnamefont {M.}~\bibnamefont
			{McCormack}}, \bibinfo {author} {\bibfnamefont {M.}~\bibnamefont
			{Linkmann}},\ and\ \bibinfo {author} {\bibfnamefont {O.}~\bibnamefont
			{Shishkina}},\ }\bibfield  {title} {\bibinfo {title} {{Unifying heat
				transport model for the transition between buoyancy-dominated and
				Lorentz-force-dominated regimes in quasistatic magnetoconvection}},\
	}\href@noop {} {\bibfield  {journal} {\bibinfo  {journal} {J. Fluid Mech.}\
		}\textbf {\bibinfo {volume} {980}},\ \bibinfo {pages} {R3} (\bibinfo {year}
		{2024})}\BibitemShut {NoStop}%
	\bibitem [{\citenamefont {Xu}\ \emph {et~al.}(2023)\citenamefont {Xu},
		\citenamefont {Horn},\ and\ \citenamefont {Aurnou}}]{Xu:PRF2023}%
	\BibitemOpen
	\bibfield  {author} {\bibinfo {author} {\bibfnamefont {Y.}~\bibnamefont
			{Xu}}, \bibinfo {author} {\bibfnamefont {S.}~\bibnamefont {Horn}},\ and\
		\bibinfo {author} {\bibfnamefont {J.~M.}\ \bibnamefont {Aurnou}},\ }\bibfield
	{title} {\bibinfo {title} {{Transition from wall modes to multimodality in
				liquid gallium magnetoconvection}},\ }\href@noop {} {\bibfield  {journal}
		{\bibinfo  {journal} {Phys. Rev. Fluids}\ }\textbf {\bibinfo {volume} {8}},\
		\bibinfo {pages} {103503} (\bibinfo {year} {2023})}\BibitemShut {NoStop}%
	\bibitem [{\citenamefont {Cresswell}\ \emph {et~al.}(2023)\citenamefont
		{Cresswell}, \citenamefont {Anders}, \citenamefont {Brown}, \citenamefont
		{Oishi},\ and\ \citenamefont {Vasil}}]{Cresswell:PRF2023}%
	\BibitemOpen
	\bibfield  {author} {\bibinfo {author} {\bibfnamefont {I.~G.}\ \bibnamefont
			{Cresswell}}, \bibinfo {author} {\bibfnamefont {E.~H.}\ \bibnamefont
			{Anders}}, \bibinfo {author} {\bibfnamefont {B.~P.}\ \bibnamefont {Brown}},
		\bibinfo {author} {\bibfnamefont {J.~S.}\ \bibnamefont {Oishi}},\ and\
		\bibinfo {author} {\bibfnamefont {G.~M.}\ \bibnamefont {Vasil}},\ }\bibfield
	{title} {\bibinfo {title} {{Force balances in strong-field magnetoconvection
				simulations}},\ }\href@noop {} {\bibfield  {journal} {\bibinfo  {journal}
			{Phys. Rev. Fluids}\ }\textbf {\bibinfo {volume} {8}},\ \bibinfo {pages}
		{093503} (\bibinfo {year} {2023})}\BibitemShut {NoStop}%
	\bibitem [{\citenamefont {Krasnov}\ \emph {et~al.}(2011)\citenamefont
		{Krasnov}, \citenamefont {Zikanov},\ and\ \citenamefont
		{Boeck}}]{Krasnov:CAF2011}%
	\BibitemOpen
	\bibfield  {author} {\bibinfo {author} {\bibfnamefont {D.}~\bibnamefont
			{Krasnov}}, \bibinfo {author} {\bibfnamefont {O.}~\bibnamefont {Zikanov}},\
		and\ \bibinfo {author} {\bibfnamefont {T.}~\bibnamefont {Boeck}},\ }\bibfield
	{title} {\bibinfo {title} {{Comparative study of finite difference
				approaches in simulation of magnetohydrodynamic turbulence at low magnetic
				Reynolds number}},\ }\href@noop {} {\bibfield  {journal} {\bibinfo  {journal}
			{Comput. Fluids}\ }\textbf {\bibinfo {volume} {50}},\ \bibinfo {pages} {46}
		(\bibinfo {year} {2011})}\BibitemShut {NoStop}%
	\bibitem [{\citenamefont {Krasnov}\ \emph {et~al.}(2023)\citenamefont
		{Krasnov}, \citenamefont {Akhtari}, \citenamefont {Zikanov},\ and\
		\citenamefont {Schumacher}}]{Krasnov:JCP2023}%
	\BibitemOpen
	\bibfield  {author} {\bibinfo {author} {\bibfnamefont {D.}~\bibnamefont
			{Krasnov}}, \bibinfo {author} {\bibfnamefont {A.}~\bibnamefont {Akhtari}},
		\bibinfo {author} {\bibfnamefont {O.}~\bibnamefont {Zikanov}},\ and\ \bibinfo
		{author} {\bibfnamefont {J.}~\bibnamefont {Schumacher}},\ }\bibfield  {title}
	{\bibinfo {title} {{Tensor-product-Thomas elliptic solver for liquid-metal
				magnetohydrodynamics}},\ }\href@noop {} {\bibfield  {journal} {\bibinfo
			{journal} {J. Comput. Phys.}\ }\textbf {\bibinfo {volume} {474}},\ \bibinfo
		{pages} {111784} (\bibinfo {year} {2023})}\BibitemShut {NoStop}%
	\bibitem [{\citenamefont {Nath}\ \emph {et~al.}(2016)\citenamefont {Nath},
		\citenamefont {Pandey}, \citenamefont {Kumar},\ and\ \citenamefont
		{Verma}}]{Nath:PRF2016}%
	\BibitemOpen
	\bibfield  {author} {\bibinfo {author} {\bibfnamefont {D.}~\bibnamefont
			{Nath}}, \bibinfo {author} {\bibfnamefont {A.}~\bibnamefont {Pandey}},
		\bibinfo {author} {\bibfnamefont {A.}~\bibnamefont {Kumar}},\ and\ \bibinfo
		{author} {\bibfnamefont {M.~K.}\ \bibnamefont {Verma}},\ }\bibfield  {title}
	{\bibinfo {title} {{Near isotropic behavior of turbulent thermal
				convection}},\ }\href@noop {} {\bibfield  {journal} {\bibinfo  {journal}
			{Phys. Rev. Fluids}\ }\textbf {\bibinfo {volume} {1}},\ \bibinfo {pages}
		{064302} (\bibinfo {year} {2016})}\BibitemShut {NoStop}%
	\bibitem [{\citenamefont {Lim}\ \emph {et~al.}(2019)\citenamefont {Lim},
		\citenamefont {Chong}, \citenamefont {Ding},\ and\ \citenamefont
		{Xia}}]{Lim:JFM2019}%
	\BibitemOpen
	\bibfield  {author} {\bibinfo {author} {\bibfnamefont {Z.~L.}\ \bibnamefont
			{Lim}}, \bibinfo {author} {\bibfnamefont {K.~L.}\ \bibnamefont {Chong}},
		\bibinfo {author} {\bibfnamefont {G.-Y.}\ \bibnamefont {Ding}},\ and\
		\bibinfo {author} {\bibfnamefont {K.-Q.}\ \bibnamefont {Xia}},\ }\bibfield
	{title} {\bibinfo {title} {{Quasistatic magnetoconvection: heat transport
				enhancement and boundary layer crossing}},\ }\href@noop {} {\bibfield
		{journal} {\bibinfo  {journal} {J. Fluid Mech,}\ }\textbf {\bibinfo {volume}
			{870}},\ \bibinfo {pages} {519} (\bibinfo {year} {2019})}\BibitemShut
	{NoStop}%
	\bibitem [{\citenamefont {Breuer}\ \emph {et~al.}(2004)\citenamefont {Breuer},
		\citenamefont {Wessling}, \citenamefont {Schmalzl},\ and\ \citenamefont
		{Hansen}}]{Breuer:PRE2004}%
	\BibitemOpen
	\bibfield  {author} {\bibinfo {author} {\bibfnamefont {M.}~\bibnamefont
			{Breuer}}, \bibinfo {author} {\bibfnamefont {S.}~\bibnamefont {Wessling}},
		\bibinfo {author} {\bibfnamefont {J.}~\bibnamefont {Schmalzl}},\ and\
		\bibinfo {author} {\bibfnamefont {U.}~\bibnamefont {Hansen}},\ }\bibfield
	{title} {\bibinfo {title} {{Effect of inertia in Rayleigh-B{\'e}nard
				convection}},\ }\href@noop {} {\bibfield  {journal} {\bibinfo  {journal}
			{Phys. Rev. E}\ }\textbf {\bibinfo {volume} {69}},\ \bibinfo {pages} {026302}
		(\bibinfo {year} {2004})}\BibitemShut {NoStop}%
	\bibitem [{\citenamefont {Scheel}\ \emph {et~al.}(2012)\citenamefont {Scheel},
		\citenamefont {Kim},\ and\ \citenamefont {White}}]{Scheel:JFM2012}%
	\BibitemOpen
	\bibfield  {author} {\bibinfo {author} {\bibfnamefont {J.~D.}\ \bibnamefont
			{Scheel}}, \bibinfo {author} {\bibfnamefont {E.}~\bibnamefont {Kim}},\ and\
		\bibinfo {author} {\bibfnamefont {K.~R.}\ \bibnamefont {White}},\ }\bibfield
	{title} {\bibinfo {title} {{Thermal and viscous boundary layers in turbulent
				Rayleigh{\textendash}B{\'e}nard convection}},\ }\href@noop {} {\bibfield
		{journal} {\bibinfo  {journal} {J. Fluid Mech.}\ }\textbf {\bibinfo {volume}
			{711}},\ \bibinfo {pages} {281} (\bibinfo {year} {2012})}\BibitemShut
	{NoStop}%
	\bibitem [{\citenamefont {Shi}\ \emph {et~al.}(2012)\citenamefont {Shi},
		\citenamefont {Emran},\ and\ \citenamefont {Schumacher}}]{Shi:JFM2012}%
	\BibitemOpen
	\bibfield  {author} {\bibinfo {author} {\bibfnamefont {N.}~\bibnamefont
			{Shi}}, \bibinfo {author} {\bibfnamefont {M.~S.}\ \bibnamefont {Emran}},\
		and\ \bibinfo {author} {\bibfnamefont {J.}~\bibnamefont {Schumacher}},\
	}\bibfield  {title} {\bibinfo {title} {{Boundary layer structure in turbulent
				Rayleigh{\textendash}B{\'e}nard convection}},\ }\href@noop {} {\bibfield
		{journal} {\bibinfo  {journal} {J. Fluid Mech.}\ }\textbf {\bibinfo {volume}
			{706}},\ \bibinfo {pages} {5} (\bibinfo {year} {2012})}\BibitemShut {NoStop}%
	\bibitem [{\citenamefont {Bhattacharya}\ \emph {et~al.}(2021)\citenamefont
		{Bhattacharya}, \citenamefont {Verma},\ and\ \citenamefont
		{Samtaney}}]{Bhattacharya:PF2020}%
	\BibitemOpen
	\bibfield  {author} {\bibinfo {author} {\bibfnamefont {S.}~\bibnamefont
			{Bhattacharya}}, \bibinfo {author} {\bibfnamefont {M.~K.}\ \bibnamefont
			{Verma}},\ and\ \bibinfo {author} {\bibfnamefont {R.}~\bibnamefont
			{Samtaney}},\ }\bibfield  {title} {\bibinfo {title} {{Revisiting Reynolds and
				Nusselt numbers in turbulent thermal convection}},\ }\href@noop {} {\bibfield
		{journal} {\bibinfo  {journal} {Phys. Fluids}\ }\textbf {\bibinfo {volume}
			{33}},\ \bibinfo {pages} {015113} (\bibinfo {year} {2021})}\BibitemShut
	{NoStop}%
	\bibitem [{\citenamefont {Verma}\ \emph {et~al.}(2017)\citenamefont {Verma},
		\citenamefont {Kumar},\ and\ \citenamefont {Pandey}}]{Verma:NJP2017}%
	\BibitemOpen
	\bibfield  {author} {\bibinfo {author} {\bibfnamefont {M.~K.}\ \bibnamefont
			{Verma}}, \bibinfo {author} {\bibfnamefont {A.}~\bibnamefont {Kumar}},\ and\
		\bibinfo {author} {\bibfnamefont {A.}~\bibnamefont {Pandey}},\ }\bibfield
	{title} {\bibinfo {title} {{Phenomenology of buoyancy-driven turbulence:
				recent results}},\ }\href@noop {} {\bibfield  {journal} {\bibinfo  {journal}
			{New J. Phys.}\ }\textbf {\bibinfo {volume} {19}},\ \bibinfo {pages} {025012}
		(\bibinfo {year} {2017})}\BibitemShut {NoStop}%
	\bibitem [{\citenamefont {Tiwari}\ \emph {et~al.}(2025)\citenamefont {Tiwari},
		\citenamefont {Sharma},\ and\ \citenamefont {Verma}}]{Tiwari:PNAS2025}%
	\BibitemOpen
	\bibfield  {author} {\bibinfo {author} {\bibfnamefont {H.}~\bibnamefont
			{Tiwari}}, \bibinfo {author} {\bibfnamefont {L.}~\bibnamefont {Sharma}},\
		and\ \bibinfo {author} {\bibfnamefont {M.~K.}\ \bibnamefont {Verma}},\
	}\bibfield  {title} {\bibinfo {title} {{On the absence of the ultimate regime
				in turbulent thermal convection}},\ }\href@noop {} {\bibfield  {journal}
		{\bibinfo  {journal} {Proc. Natl. Acad. Sci. U.S.A.}\ }\textbf {\bibinfo
			{volume} {122}},\ \bibinfo {pages} {e2513474122} (\bibinfo {year}
		{2025})}\BibitemShut {NoStop}%
	\bibitem [{\citenamefont {Shraiman}\ and\ \citenamefont
		{Siggia}(1990)}]{Shraiman:PRA1990}%
	\BibitemOpen
	\bibfield  {author} {\bibinfo {author} {\bibfnamefont {B.~I.}\ \bibnamefont
			{Shraiman}}\ and\ \bibinfo {author} {\bibfnamefont {E.~D.}\ \bibnamefont
			{Siggia}},\ }\bibfield  {title} {\bibinfo {title} {{Heat transport in
				high-Rayleigh-number convection}},\ }\href@noop {} {\bibfield  {journal}
		{\bibinfo  {journal} {Phys. Rev. A}\ }\textbf {\bibinfo {volume} {42}},\
		\bibinfo {pages} {3650} (\bibinfo {year} {1990})}\BibitemShut {NoStop}%
	\bibitem [{\citenamefont {Z{\"u}rner}\ \emph {et~al.}(2016)\citenamefont
		{Z{\"u}rner}, \citenamefont {Liu}, \citenamefont {Krasnov},\ and\
		\citenamefont {Schumacher}}]{Zurner:PRE2016}%
	\BibitemOpen
	\bibfield  {author} {\bibinfo {author} {\bibfnamefont {T.}~\bibnamefont
			{Z{\"u}rner}}, \bibinfo {author} {\bibfnamefont {W.}~\bibnamefont {Liu}},
		\bibinfo {author} {\bibfnamefont {D.}~\bibnamefont {Krasnov}},\ and\ \bibinfo
		{author} {\bibfnamefont {J.}~\bibnamefont {Schumacher}},\ }\bibfield  {title}
	{\bibinfo {title} {{Heat and momentum transfer for magnetoconvection in a
				vertical external magnetic field}},\ }\href@noop {} {\bibfield  {journal}
		{\bibinfo  {journal} {Phys. Rev. E}\ }\textbf {\bibinfo {volume} {94}},\
		\bibinfo {pages} {043108} (\bibinfo {year} {2016})}\BibitemShut {NoStop}%
	\bibitem [{\citenamefont {Z{\"u}rner}(2020)}]{Zurner:PF2020}%
	\BibitemOpen
	\bibfield  {author} {\bibinfo {author} {\bibfnamefont {T.}~\bibnamefont
			{Z{\"u}rner}},\ }\bibfield  {title} {\bibinfo {title} {{Refined mean field
				model of heat and momentum transfer in magnetoconvection}},\ }\href@noop {}
	{\bibfield  {journal} {\bibinfo  {journal} {Phys. Fluids}\ }\textbf {\bibinfo
			{volume} {32}},\ \bibinfo {pages} {107101} (\bibinfo {year}
		{2020})}\BibitemShut {NoStop}%
	\bibitem [{\citenamefont {Grossmann}\ and\ \citenamefont
		{Lohse}(2001)}]{Grossmann:PRL2001}%
	\BibitemOpen
	\bibfield  {author} {\bibinfo {author} {\bibfnamefont {S.}~\bibnamefont
			{Grossmann}}\ and\ \bibinfo {author} {\bibfnamefont {D.}~\bibnamefont
			{Lohse}},\ }\bibfield  {title} {\bibinfo {title} {{Thermal Convection for
				Large Prandtl Numbers}},\ }\href@noop {} {\bibfield  {journal} {\bibinfo
			{journal} {Phys. Rev. Lett.}\ }\textbf {\bibinfo {volume} {86}},\ \bibinfo
		{pages} {3316} (\bibinfo {year} {2001})}\BibitemShut {NoStop}%
	\bibitem [{\citenamefont {Bhattacharya}\ \emph {et~al.}(2018)\citenamefont
		{Bhattacharya}, \citenamefont {Pandey}, \citenamefont {Kumar},\ and\
		\citenamefont {Verma}}]{Bhattacharya:PF2018}%
	\BibitemOpen
	\bibfield  {author} {\bibinfo {author} {\bibfnamefont {S.}~\bibnamefont
			{Bhattacharya}}, \bibinfo {author} {\bibfnamefont {A.}~\bibnamefont
			{Pandey}}, \bibinfo {author} {\bibfnamefont {A.}~\bibnamefont {Kumar}},\ and\
		\bibinfo {author} {\bibfnamefont {M.~K.}\ \bibnamefont {Verma}},\ }\bibfield
	{title} {\bibinfo {title} {{Complexity of viscous dissipation in turbulent
				thermal convection}},\ }\href@noop {} {\bibfield  {journal} {\bibinfo
			{journal} {Phys. Fluids}\ }\textbf {\bibinfo {volume} {30}},\ \bibinfo
		{pages} {031702} (\bibinfo {year} {2018})}\BibitemShut {NoStop}%
	\bibitem [{\citenamefont {Bhattacharya}\ \emph {et~al.}(2019)\citenamefont
		{Bhattacharya}, \citenamefont {Samtaney},\ and\ \citenamefont
		{Verma}}]{Bhattacharya:PF2019}%
	\BibitemOpen
	\bibfield  {author} {\bibinfo {author} {\bibfnamefont {S.}~\bibnamefont
			{Bhattacharya}}, \bibinfo {author} {\bibfnamefont {R.}~\bibnamefont
			{Samtaney}},\ and\ \bibinfo {author} {\bibfnamefont {M.~K.}\ \bibnamefont
			{Verma}},\ }\bibfield  {title} {\bibinfo {title} {{Scaling and spatial
				intermittency of thermal dissipation in turbulent convection}},\ }\href@noop
	{} {\bibfield  {journal} {\bibinfo  {journal} {Phys. Fluids}\ ,\ \bibinfo
			{pages} {1}} (\bibinfo {year} {2019})}\BibitemShut {NoStop}%
	\bibitem [{\citenamefont {Kolmogorov}(1962)}]{Kolmogorov:JFM1963}%
	\BibitemOpen
	\bibfield  {author} {\bibinfo {author} {\bibfnamefont {A.~N.}\ \bibnamefont
			{Kolmogorov}},\ }\bibfield  {title} {\bibinfo {title} {{A refinement of
				previous hypotheses concerning the local structure of turbulence in a viscous
				incompressible fluid at high Reynolds number}},\ }\href@noop {} {\bibfield
		{journal} {\bibinfo  {journal} {J. Fluid Mech.}\ }\textbf {\bibinfo {volume}
			{13}},\ \bibinfo {pages} {82} (\bibinfo {year} {1962})}\BibitemShut {NoStop}%
	\bibitem [{\citenamefont {Schumacher}\ and\ \citenamefont
		{Sreenivasan}(2005)}]{Schumacher:PF2005}%
	\BibitemOpen
	\bibfield  {author} {\bibinfo {author} {\bibfnamefont {J.}~\bibnamefont
			{Schumacher}}\ and\ \bibinfo {author} {\bibfnamefont {K.~R.}\ \bibnamefont
			{Sreenivasan}},\ }\bibfield  {title} {\bibinfo {title} {{Statistics and
				geometry of passive scalars in turbulence}},\ }\href@noop {} {\bibfield
		{journal} {\bibinfo  {journal} {Phys. Fluids}\ }\textbf {\bibinfo {volume}
			{17}},\ \bibinfo {pages} {125107} (\bibinfo {year} {2005})}\BibitemShut
	{NoStop}%
	\bibitem [{\citenamefont {Emran}\ and\ \citenamefont
		{Schumacher}(2008)}]{Emran:JFM2008}%
	\BibitemOpen
	\bibfield  {author} {\bibinfo {author} {\bibfnamefont {M.~S.}\ \bibnamefont
			{Emran}}\ and\ \bibinfo {author} {\bibfnamefont {J.}~\bibnamefont
			{Schumacher}},\ }\bibfield  {title} {\bibinfo {title} {{Fine-scale statistics
				of temperature and its derivatives in convective turbulence}},\ }\href@noop
	{} {\bibfield  {journal} {\bibinfo  {journal} {J. Fluid Mech.}\ }\textbf
		{\bibinfo {volume} {611}},\ \bibinfo {pages} {13} (\bibinfo {year}
		{2008})}\BibitemShut {NoStop}%
	\bibitem [{\citenamefont {Kaczorowski}\ and\ \citenamefont
		{Wagner}(2008)}]{Kaczorowski:JFM2008}%
	\BibitemOpen
	\bibfield  {author} {\bibinfo {author} {\bibfnamefont {M.}~\bibnamefont
			{Kaczorowski}}\ and\ \bibinfo {author} {\bibfnamefont {C.}~\bibnamefont
			{Wagner}},\ }\bibfield  {title} {\bibinfo {title} {{Analysis of the thermal
				plumes in turbulent Rayleigh{\textendash}B{\'e}nard convection based on
				well-resolved numerical simulations}},\ }\href@noop {} {\bibfield  {journal}
		{\bibinfo  {journal} {J. Fluid Mech.}\ }\textbf {\bibinfo {volume} {618}},\
		\bibinfo {pages} {89} (\bibinfo {year} {2008})}\BibitemShut {NoStop}%
	\bibitem [{\citenamefont {Zhang}\ \emph {et~al.}(2017)\citenamefont {Zhang},
		\citenamefont {Zhou},\ and\ \citenamefont {Sun}}]{Zhang:JFM2017}%
	\BibitemOpen
	\bibfield  {author} {\bibinfo {author} {\bibfnamefont {Y.}~\bibnamefont
			{Zhang}}, \bibinfo {author} {\bibfnamefont {Q.}~\bibnamefont {Zhou}},\ and\
		\bibinfo {author} {\bibfnamefont {C.}~\bibnamefont {Sun}},\ }\bibfield
	{title} {\bibinfo {title} {{Statistics of kinetic and thermal energy
				dissipation rates in two-dimensional turbulent
				Rayleigh{\textendash}B{\'e}nard convection}},\ }\href@noop {} {\bibfield
		{journal} {\bibinfo  {journal} {J. Fluid Mech.}\ }\textbf {\bibinfo {volume}
			{814}},\ \bibinfo {pages} {165} (\bibinfo {year} {2017})}\BibitemShut
	{NoStop}%
\end{thebibliography}

\end{document}